\documentclass[12pt,twoside,a4paper]{article}
\pdfoutput=1
\usepackage{graphicx}
\usepackage{url}
\usepackage{color}
\usepackage{cite}

\def\lsim{\:\raisebox{-0.5ex}{$\stackrel{\textstyle<}{\sim}$}\:}

\begin{document}
\thispagestyle{empty} 

\title{
\vskip-3cm
{\baselineskip14pt
\centerline{\normalsize DESY 18-154 \hfill ISSN 0418--9833}
\centerline{\normalsize MITP/18--085 \hfill} 
\centerline{\normalsize September 2018 \hfill}} 
\vskip1.5cm
\boldmath
{\bf $b$-Hadron production in the}
\\
{\bf general-mass variable-flavour-number scheme}
\\
{\bf and LHC data}
\unboldmath
\author{
G.~Kramer$^1$, 
and H.~Spiesberger$^2$
\vspace{2mm} \\
\normalsize{
  $^1$ II. Institut f\"ur Theoretische
  Physik, Universit\"at Hamburg,
}\\ 
\normalsize{
  Luruper Chaussee 149, D-22761 Hamburg, Germany
} \vspace{2mm}
\\
  {\normalsize 
  $^2$ PRISMA Cluster of Excellence, Institut f\"ur Physik,}
  \\
  {\normalsize 
  Johannes Gutenberg-Universit\"at, 55099 Mainz, Germany,}
  \\
  {\normalsize 
  and Centre for Theoretical and Mathematical Physics and 
  Department of Physics,}
  \\
  {\normalsize 
  University of Cape Town, Rondebosch 7700, South Africa}
\vspace{2mm} 
\\
}}

\maketitle
\begin{abstract}
\medskip
\noindent
We study inclusive $b$-hadron production in $pp$ collisions 
at the LHC at different center-of-mass energies and compare 
with experimental data from the LHCb and CMS collaborations. 
Our predictions for cross sections differential in the 
transverse momentum and (pseudo-)rapidity agree with data 
within uncertainties due to renormalization scale variations. 
A small tension is found if data and theory predictions 
are compared for cross section ratios at different 
center-of-mass energies. 
\end{abstract}

\clearpage

\section{Introduction}

The investigation of inclusive production of hadrons 
containing $b$ quarks is particularly important to test 
quantum chromodynamics (QCD). The predictions in the 
framework of perturbative QCD are based on the factorization 
approach. Cross sections are calculated as a convolution of 
three basic parts: the parton distribution functions (PDF) 
describing the parton content of the initial hadronic state, 
the partonic hard scattering cross sections computed as a 
perturbative series in powers of the strong coupling constant, 
and the fragmentation functions (FF), which describe the 
production yield and the momentum distribution for a specified 
$b$ hadron in a parton. Since the $b$-quark mass is large 
and can often not be neglected compared with the transverse 
momentum, the cross section for $b$-hadron production depends 
on several large scales, which makes predictions very 
challenging. 

In the past, measurements of inclusive $b$-hadron production 
and the corresponding QCD calculations were done first of 
all for the $B$-mesons, i.e.\ $B^{\pm}$, $B^0$, 
$\overline{B}^0$, $B_s^0$, $\overline{B}^0_s$, but also 
$\Lambda_b^0$ and other $b$-baryons have been considered. 
Data for $p\bar{p}$ collisions at $\sqrt{S} = 1.96$~TeV 
have been obtained at the FNAL Tevatron Collider 
\cite{Acosta:2004yw,Abulencia:2006ps} and for $pp$ 
collisions at $\sqrt{S} = 5$, $7$, $8$ and $13$~TeV 
at the CERN Large Hadron Collider (LHC) by the ATLAS, 
CMS and LHCb collaborations 
\cite{ATLAS:2013cia,Khachatryan:2011mk,Chatrchyan:2011pw, 
Chatrchyan:2011vh,Aaij:2012jd,Khachatryan:2016csy}. First 
measurements of the production cross sections of 
$\Lambda^0_b$ baryons have been performed by the CMS 
collaboration at the LHC \cite{Chatrchyan:2012xg} at 
$\sqrt{S} = 7$ TeV and by the LHCb collaboration for
$\sqrt{S} = 7$ and $8$ TeV \cite{Aaij:2015fea}.

Almost all of these data have been compared with 
next-to-leading order (NLO) QCD predictions based on 
the so-called FONLL approach \cite{Cacciari:2012ny}. 
Data of the CMS, LHCb and ATLAS collaborations have 
also been compared with predictions obtained in the 
general-mass variable-flavour-number scheme (GM-VFNS) 
\cite{Kniehl:2011bk,Kniehl:2015fla}. The GM-VFNS 
\cite{Kniehl:2004fy,Kniehl:2005mk} (see also 
\cite{Helenius:2018uul} for a more recent implementation 
of the GM-VFNS) is similar to the FONLL scheme but 
contains different assumptions concerning 
fragmentation functions, and the transition to the 
fixed-flavor-number-scheme (FFNS) in the low transverse 
momentum, $p_T$, region is treated in a different way. 
All comparisons between experimental data and theoretical predictions, whether based on the FONLL approach or on 
the GM-VFNS, show reasonable agreement within experimental 
uncertainties and taking into account the so-called
theoretical error which is estimated by a variation 
of the factorization and renormalization scale parameters 
or the heavy quark masses.

An exception from agreement between experimental data and 
theoretical predictions had originally been reported by 
the LHCb collaboration \cite{Aaij:2016avz,CERN-Courier56}. 
Their primary finding that the ratio $R_{13/7}$ of cross 
sections $d\sigma/d\eta$ as a function of the pseudorapidity 
$\eta$ in the region $2 < \eta < 5$ for $\sqrt{S} = 13$ 
and $\sqrt{S} = 7$~TeV did not agree with corresponding 
FONLL predictions \cite {Cacciari:2015fta} was corrected 
later \cite{Aaij-Err:2016avz} after a mistake in their 
detector simulation was found which changed the data 
for the 13~TeV $b$-quark production cross section. In the 
LHCb analysis, the observed semileptonic decays of $b$ 
hadrons was corrected in such a way that the extracted 
cross section can be interpreted as a measurement of the 
inclusive $b$-quark production, not just for one specific 
$B$-meson species. 

Other measurements of $b$-quark production show some 
discrepancy between data and theoretical predictions, 
although with smaller significance than what was reported 
for the now corrected LHCb data. Among them we count the 
data from the CMS collaboration for inclusive $B^+$ 
production \cite{Khachatryan:2016csy}. Here, $B^+$ 
mesons were identified by their decay into $J/\psi K^+$ 
final states at the energies $\sqrt{S}$ = 7 and 13~TeV. 
Differential cross sections were determined both as a 
function of $p_T$ (integrated over rapidity ranges $|y| 
< 1.45$ and $|y| < 2.1$), as well as a function of $|y|$ 
(integrated over $p_T$ in the ranges $10 < p_T < 100$~GeV 
and $17 < p_T < 100$~GeV). A discrepancy was observed in 
a comparison with predictions obtained in the FONLL approach. 

The purpose of our present work is to present results 
for $b$-hadron production cross sections and the cross 
section ratios $R_{13/7}$ in $pp$ collisions at the LHC 
in the framework of the GM-VFNS. This framework is 
essentially the conventional NLO QCD parton-model 
approach supplemented with heavy-quark finite mass 
effects intended to improve the description at small 
and medium transverse momenta. First, we shall apply 
the GM-VFNS approach for the calculation of the NLO 
single-inclusive cross sections. Second, we shall 
investigate the dependence of the cross section ratio 
$R_{13/7}$ on the assumed input PDF. We will show that 
the cross section ratios have much smaller theoretical 
uncertainties than the differential cross sections 
themselves. 

The outline of the paper is as follows. In the next section 
we introduce some details of the calculation, describe our 
choice of the proton PDFs and the fragmentation functions, 
and discuss how rapidity and pseudorapidity distributions 
are related. In Sect.~\ref{Sec:complhcb} we collect our 
results for inclusive $b$-hadron cross sections 
$d\sigma/d\eta$ at 7 and 13~TeV and compare with the LHCb 
data. A similar comparison with CMS data, including also 
the $p_T$ distribution, is performed in Sect.~\ref{Sec:compcms} 
at 7 and 13~TeV in the central rapidity region. In addition, 
we discuss theory predictions and experimental results 
for $J/\Psi$ production in Sect.~\ref{Sec:compjpsi}. Our 
conclusions with some outlook are presented in 
Sect.\ref{Sec:conclusion}.

\section{Setup, input PDFs and FFs}
\label{Sec:setup}

The theoretical description of the GM-VFNS approach as well 
as the technical details of its implementation have been 
presented previously in Refs.~\cite{Kniehl:2004fy,Kniehl:2005mk}. 
Here we describe only the input required for the numerical 
evaluations discussed below. 

As a default we use the proton PDF set CT14~\cite{Dulat:2015mca} 
at NLO as implemented in the LHAPDF library~\cite{Buckley:2014ana}. 
To study the sensitivity on the PDF input we shall use two 
approaches. First we will use three alternative PDF sets: 
(i) HERA2.0~\cite{Abramowicz:2015mha}, 
(ii) MMHT~\cite{Harland-Lang:2014zoa} and 
(iii) NNPDF3.0~\cite{Ball:2014uwa}. All these PDF sets 
are NLO parametrizations, the last two of them are obtained 
from global fits to essentially the same experimental data 
as CT14, while the set HERA2.0 is based mainly on cross 
section data for deep inelastic scattering at HERA. 
Secondly, for the CT14 parametrization we will also study 
uncertainties obtained from variations of parameter 
eigenvalues. There are 56 members of the CT14 set corresponding 
to 28 pairs of eigenvalue variations. One pair is particularly 
interesting since it describes an enhanced/suppressed gluon 
distribution at very low $x$. The PDF uncertainty band is 
evaluated following the prescription given in Eq.~(5) of 
\cite{Lai:2010vv} and corresponds to $90\%$ CL. 

To describe the transition of $b$ quarks to $b$ hadrons 
we need non-perturbative FFs. We employ the $B$-meson FFs 
constructed in \cite{Kniehl:2008zza}. They are evolved 
at NLO and components for the transition from gluons 
and light quarks (including charm) to a $B$ meson are 
generated through DGLAP evolution. They were obtained by 
fitting experimental data for inclusive $b$ production 
in $e^+e^-$ annihilation taken by the ALEPH 
\cite{Heister:2001jg} and OPAL \cite{Abbiendi:2002vt} 
collaborations at CERN LEP1 and by the SLD collaboration 
\cite{Abe:1999ki,Abe:2002iq} at SLAC SLC. These data were 
all taken on the $Z$-boson resonance. Therefore 
$\alpha_s^{(n_f)}(\mu_R)$ was evaluated with $n_f = 5$ 
and the renormalization and factorization scales were 
fixed at $\mu_R = \mu_F = m_Z$. The starting scale was 
chosen to be $\mu_0 = m_b = 4.5$~GeV. Below $\mu_F = 
\mu_0$ the light-quark and gluon FFs were assumed to 
vanish. A simple power ansatz gave the best fit to the 
experimental data.

One should notice that the $B$-meson FFs of 
Ref.~\cite{Kniehl:2008zza} do not distinguish between 
different $b$-hadron final states. Both the OPAL 
\cite{Abbiendi:2002vt} and the SLD 
\cite{Abe:1999ki,Abe:2002iq} data include all 
$b$ hadrons, i.e.\ the mesons $B^{\pm}$, $B^0$, 
$\overline{B}^0$, $B_s^0$ and $\overline{B}_s^0$ as well as 
$b$ baryons, while in the ALEPH analysis \cite{Heister:2001jg} 
only final states with identified $B^{\pm}$, $B^0$ and 
$\overline{B}^0$ mesons were taken into account. Despite 
of these differences in the experimental analyses it was 
assumed in \cite{Kniehl:2008zza} that all data can be 
described by one common FF. The resulting FF fit did 
indeed not show any significant difference with either 
of the two data sets, of OPAL and SLD including all 
$b$-hadrons on the one side, or of ALEPH including only 
identified $B$-mesons on the other side. The FF was 
normalized to describe cross section data for $B^+$ and 
$B^0$-meson production. They can also be used to calculate 
the sum of all $b$-hadron states by removing the fragmentation 
fraction for the $b \to B^{\pm}$ transition, which was 
assumed as $f_u = f_d = 0.397$ in \cite{Kniehl:2008zza}.

For simplicity we shall take the initial- and final-state 
factorization scales entering the PDFs and FFs, respectively, 
to have the same value, denoted by $\mu_F$. The majority 
of data to which we are going to compare our theory 
predictions is dominated by low transverse momenta. 
For example, the LHCb data for the pseudorapidity 
distribution, $d\sigma/d\eta$, is integrated over 
all $p_T$, down to $p_T = 0$. In a previous work 
\cite{Kniehl:2015fla} we have shown that the choice 
$\mu_F = \xi_F \sqrt{p_T^2+m_b^2}$ with $\xi_F = 0.5$ 
for the factorization scale is appropriate in this case. 
This choice allows a transition to the fixed-flavour 
number scheme at finite values of the transverse 
momentum. The FFNS is the appropriate prescription for 
heavy-quark production in the range $p_T \lsim m_b$. 
In the FFNS, sub-processes with heavy quarks in the 
initial state do not contribute. With our choice of 
$\mu_F$, the scale reaches the heavy-quark threshold 
$\mu_F = m_b$ already at $p_T = 7.8$~GeV below which 
both the $b$-PDF and the FFs are zero. In 
Ref.~\cite{Kniehl:2015fla} 
we could show that with $\xi_F = 0.5$ the cross sections 
$d\sigma/dp_T$ as a function of $p_T$ as measured by LHCb 
at $\sqrt{S} = 7$~TeV \cite{Aaij:2013noa} and by CDF 
at the Tevatron~\cite{Acosta:2004yw} can be described 
satisfactorily well down to $p_T = 0$. The factor $\xi_F = 
0.5$ is not unique, but small variations like $\xi_F = 0.4$ 
and $0.6$ lead only to small changes, as shown in 
Ref.~\cite{Kniehl:2015fla}. 

We evaluate the strong coupling $\alpha_s^{(n_f)}(\mu_R)$ 
as a function of the renormalization scale $\mu_R$ at NLO 
with $\Lambda^{(4)}_{\overline{MS}} = 328$~MeV for 
$n_f = 4$ flavours. This corresponds to 
$\Lambda^{(5)}_{\overline{MS}}= 226$~MeV above the 5 flavour 
threshold at $\mu_R = m_b$. For the $b$-quark pole mass we 
use $m_b = 4.5$~GeV in accordance with the value chosen 
for the FF fit in Ref.~\cite{Kniehl:2008zza}. This value 
is also compatible, though not identical, with the 
$b$-quark thresholds in the PDF parametrizations which 
we are going to use. 

\begin{figure*}[t!]
\begin{center}
\includegraphics[width=9cm]{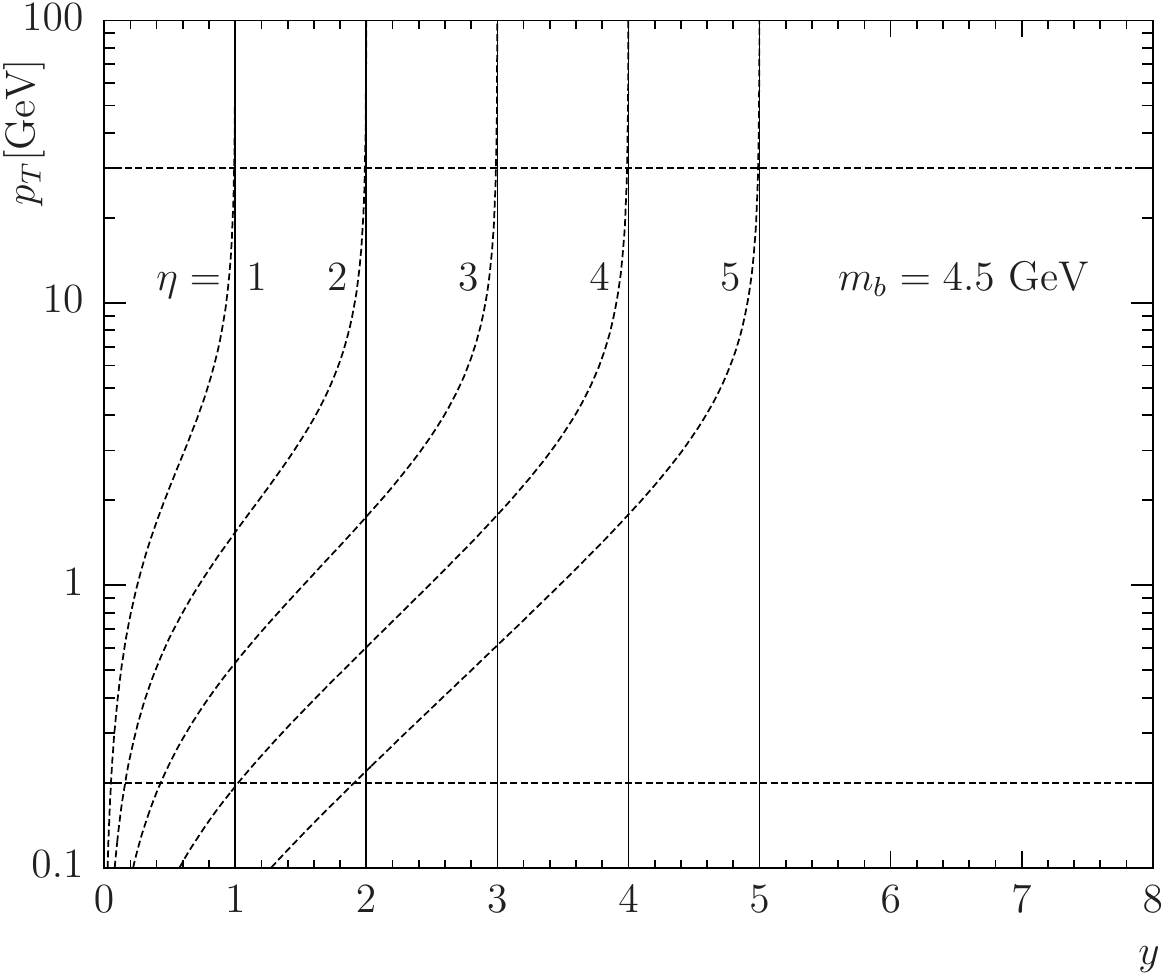}
\end{center}
\caption{
Lines of constant pseudorapidity as a function of the 
rapidity for mass $m_b=4.5$~GeV in the $p_T$ range relevant 
for LHCb data. The horizontal dashed lines indicate the 
values of $p_T = 0.2$ and $=30$~GeV.}
\label{fig:etatoy} 
\end{figure*}

We have updated the implementation of our program for the 
calculation of GM-VFNS predictions to allow integration 
with fixed pseudorapidity. Results in our previous 
publications have always been given for rapidity intervals. 
We note that the difference between rapidity and pseudorapidity 
cross sections is quite substantial for the LHCb data. 
In Fig.~\ref{fig:etatoy} we show contours of constant 
pseudorapidity in the plane of transverse momentum 
$p_T$ and rapidity $y$. The calculation is made for the 
kinematic range relevant for the LHCb experiment to be 
discussed in the next section and using $m_b = 4.5$~GeV. 
One can see that at low $p_T$, where the dominating 
contribution to the cross section $d\sigma/d\eta$ is 
found, the average rapidity is shifted to much smaller 
values compared with the pseudorapidity. It turns out 
that $d\sigma/d\eta$ is increased, compared with 
$d\sigma/dy$, by 50 to 100\,\% in the high $\eta$-bins 
of the LHCb measurements while it is almost the same at 
smaller values of the (pseudo-)rapidity. The effect on 
the cross section is not as strong as could be expected 
from Fig.~\ref{fig:etatoy} since the $\eta$-dependence 
of the differential cross section is not very strong, 
but the difference is still relevant. We also note that 
the results depend little on the exact choice of the 
particle mass in the relation between rapidity and 
pseudorapidity and our conclusions would not change 
if we had used a value corresponding to the lightest 
$b$ meson ($m_{B^0} = 5.28$~GeV) or even the heavier 
$b$ hadron ($m_{\Lambda_b} = 5.62$~GeV).

\section{Comparison with LHCb data}
\label{Sec:complhcb}

\begin{figure*}[b!]
\begin{center}
\includegraphics[width=7.8cm]{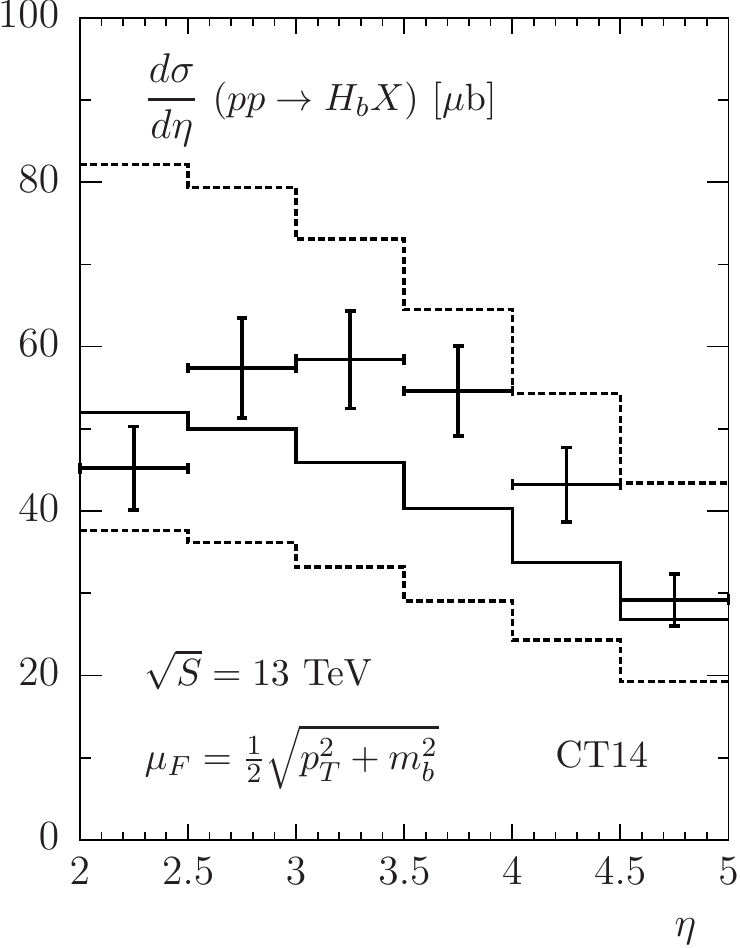}
\includegraphics[width=7.55cm]{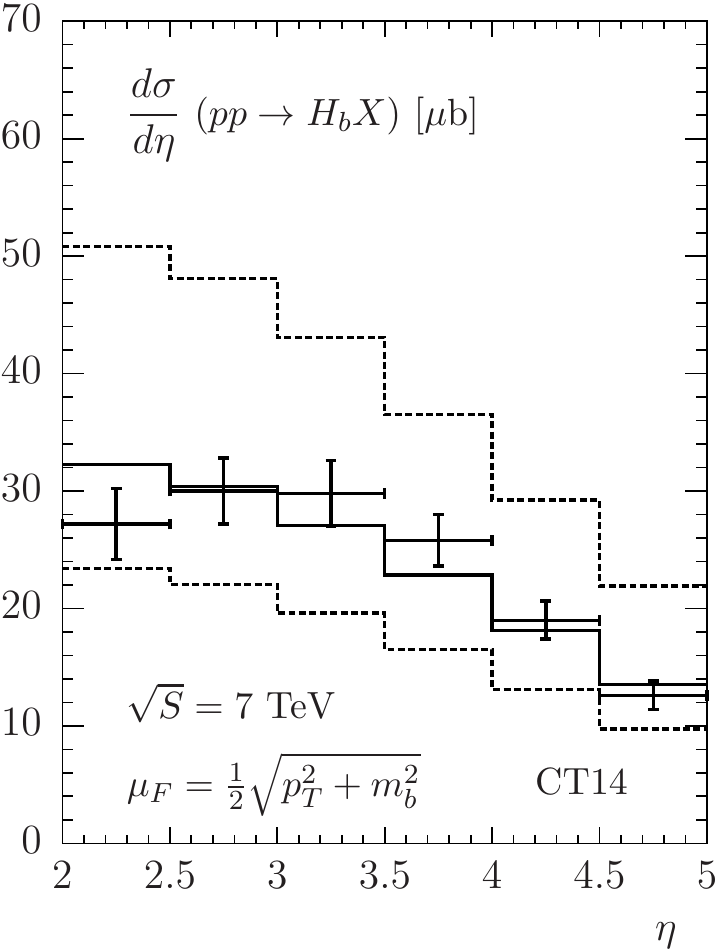}
\end{center}
\caption{
Pseudorapidity distribution for $b$-hadron production compared 
with LHCb data \cite{Aaij-Err:2016avz}, using CT14, $\sqrt{S} 
= 13$ TeV (left) and 7 TeV (right). The dashed histograms 
describe the theory uncertainty due to variations of the 
renormalization scale. The upper limit is found for $\mu_R 
= 0.5 \sqrt{p_T^2+m_b^2}$, the lower limit for $\mu_R = 
2.0 \sqrt{p_T^2+m_b^2}$. 
}
\label{fig:1} 
\end{figure*}

The LHCb collaboration has reported cross section measurements 
for $b$-hadron production in $pp$ collisions at $\sqrt{S} = 7$ 
and 13~TeV \cite{Aaij:2016avz,Aaij-Err:2016avz}. Results 
are given for the pseudorapidity dependence, $d\sigma/d\eta$, 
in 6 equal-sized bins in the range $2 < \eta < 5$. 
$b$ hadrons are identified by their semileptonic decays 
into a ground-state charmed hadron in association with a 
muon. We show these data together with our results (full 
line histograms) for the proton PDFs of CT14 
\cite{Dulat:2015mca} in Fig.~\ref{fig:1}. We have used 
the factorization scale $\mu_F = 0.5 \sqrt{p_T^2+m_b^2}$ 
with $m_b = 4.5$ GeV as in \cite{Kniehl:2015fla}. The 
renormalization scale was fixed as $\mu_R=\sqrt{p_T^2+m_b^2}$ 
and varied by a factor of two up and down to obtain an 
estimate of the theoretical uncertainty. This results in 
the dashed-line histograms in Fig.~\ref{fig:1}. The upper 
limit of the uncertainty band is found for $\mu_R = \mu_F 
= 0.5 \sqrt{p_T^2+m_b^2}$ and the lower limit for $\mu_R 
= 4 \mu_F = 2.0 \sqrt{p_T^2+m_b^2}$. We do not observe a 
discrepancy between data and theoretical results: all data 
points lie inside the theoretical band for both values of 
$\sqrt{S}$ and in all $\eta$ bins. However, the shape 
of the $\eta$-dependence at values $\eta \lsim 3$, 
where the cross section data decrease with decreasing 
$\eta$, is not visible in the theory prediction. 

\begin{figure*}[b!]
\begin{center}
\includegraphics[width=7.8cm]{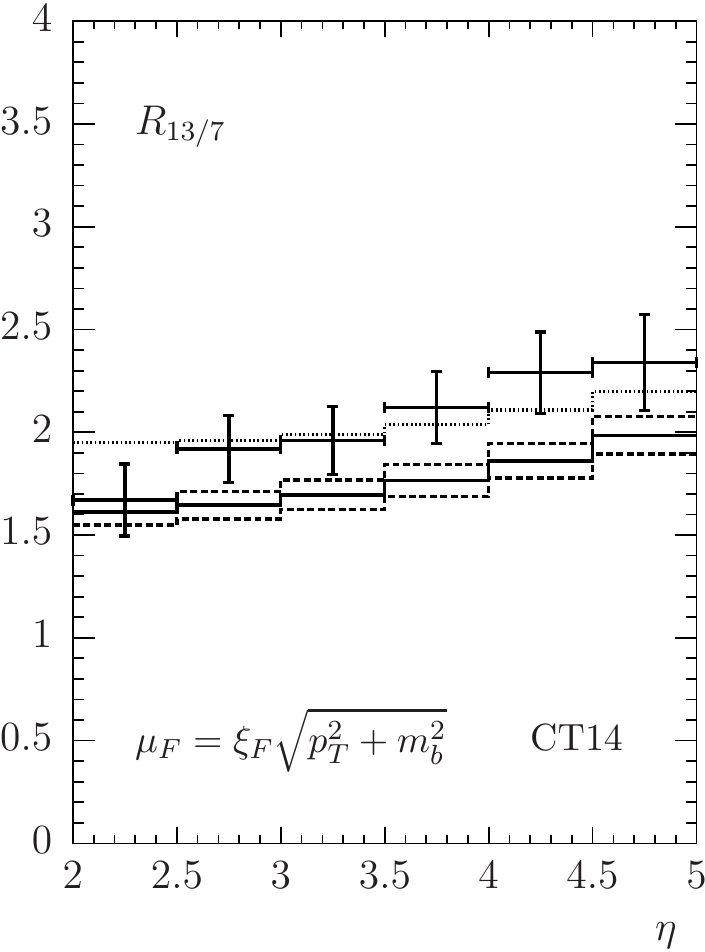}
\end{center}
\caption{
The ratio of the pseudorapidity distributions for $\sqrt{S} 
= 13$ and 7 TeV for CT14 PDFs compared with LHCb data 
\cite{Aaij-Err:2016avz}. The full and dashed histograms are 
calculated with $\xi_F=0.5$, as in Fig.~\ref{fig:1}. Its 
error band (dashed lines) shows the statistical uncertainty 
of the Monte Carlo integration. The dotted histogram is 
obtained with $\xi_F=0.7$ for the cross section at 
$\sqrt{S} = 13$ and with $\xi_F=0.5$ at $\sqrt{S} = 7$~TeV. 
}
\label{fig:2} 
\end{figure*}

In Fig.~\ref{fig:2} we show the ratio $R_{13/7}$ of the 
cross sections $d\sigma/d\eta$ for $\sqrt{S} = 13$ and 
7~TeV and compare our results with the experimental values 
in the six $\eta$-bins. The figure shows predictions for 
the CT14 PDFs (full-line histogram). The theoretical 
result varies between 1.6 and 2.0. The uncertainty due 
to the variation of the renormalization scale is almost 
completely cancelled in the ratio. In fact, our numerical 
calculation is dominated by the statistical uncertainty 
of the Monte Carlo integration -- the error band shown in 
Fig.~\ref{fig:2} (dashed-line histograms) is representing 
this numerical uncertainty.  

The cross section ratio $R_{13/7}$ is not affected by 
the large uncertainties from scale variations since one 
should use the same prescription for fixing the scales 
in the numerator and in the denominator. The theory 
uncertainties of the cross sections at the different 
center-of-mass energies can not be treated like 
experimental errors, which have to be added in 
quadrature. We note that the agreement between data 
and theory could be improved if we used different 
scale choices for the different energies. As an example, 
we show in Fig.~\ref{fig:2} how the ratio $R_{13/7}$ is 
changed if the cross section for $\sqrt{S} = 13$~TeV is 
calculated with $\mu_F = 0.7 \sqrt{p_T^2+m_b^2}$ 
instead of $\mu_F = 0.5 \sqrt{p_T^2+m_b^2}$; the 
latter choice of scale is kept for the calculation 
at $\sqrt{S} = 7$~TeV. This prescription moves the 
ratio up in such a way that the data are perfectly 
well described in the upper five bins. Only the first 
bin shows a slight discrepancy. Of course, such a 
prescription to fix the factorization scale would be 
rather ad-hoc and we have no theoretical justification 
for it, however, it is also not forbidden by theory. We do 
not attempt to introduce an additional $\eta$ dependence 
of scales, because such an approach would be even less 
well motivated.  

\begin{figure*}[b!]
\begin{center}
\includegraphics[width=7.9cm]{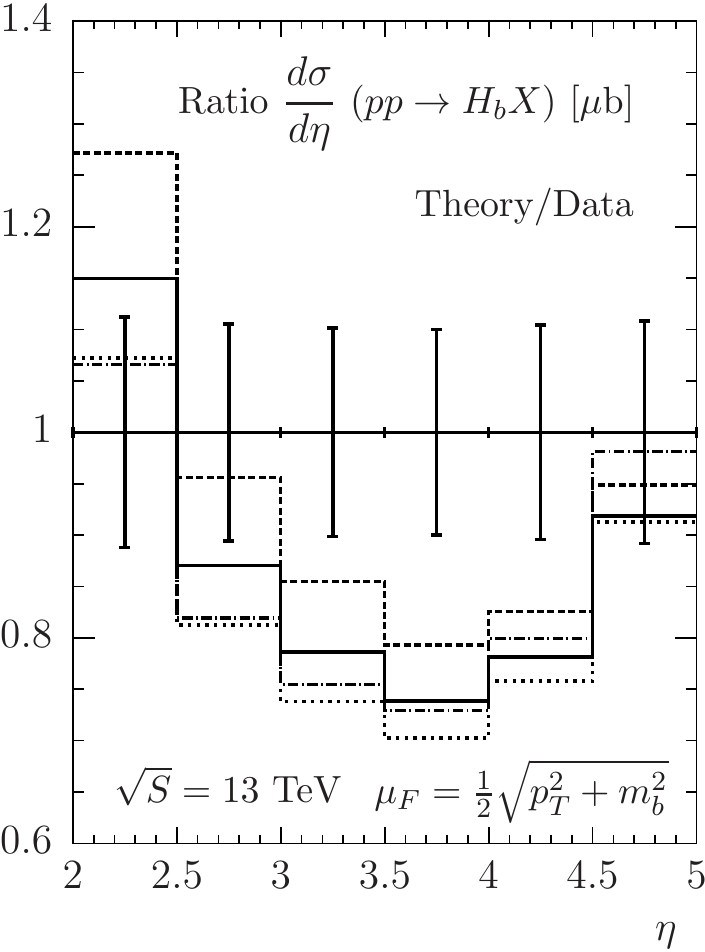}
\includegraphics[width=7.9cm]{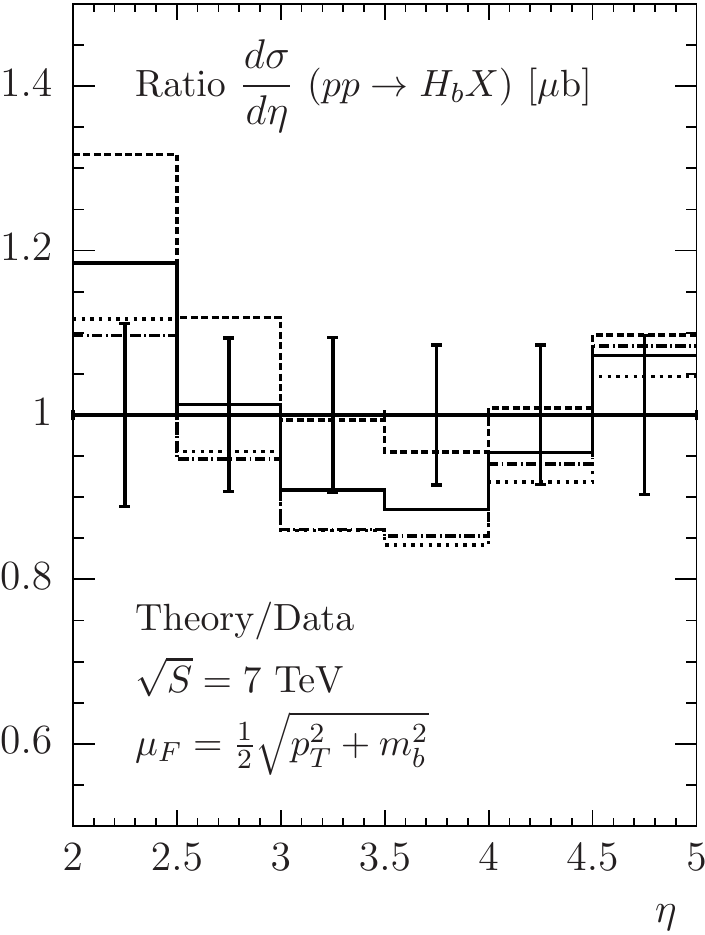}
\end{center}
\caption{
PDF uncertainties of the pseudorapidity distributions for 
$b$-hadron production. The plot shows theory predictions 
normalized to the LHCb data \cite{Aaij-Err:2016avz} for $\sqrt{S} 
= 13$~TeV (left) and 7~TeV (right). Full line: CT14, dashed 
line: HERAPDF2.0, dotted line: NNPDF3.0, dash-dotted line MMHT. 
For better visibility, scale uncertainties for the theory 
predictions are not shown. The corresponding uncertainty 
band would lie partly outside the range shown in the figure. 
}
\label{fig:3} 
\end{figure*}

It is well-known that the theory uncertainty for the 
cross sections $d\sigma/d\eta$ is dominated by far by 
scale variations. These errors cancel, however, in the 
ratio $R_{13/7}$. We therefore shall have a closer look 
at uncertainties due to the PDF input. First, in 
Fig.~\ref{fig:3} we show results where we have used also 
the PDF parametrizations HERAPDF2.0 \cite{Abramowicz:2015mha}, 
MMHT \cite{Harland-Lang:2014zoa} and NNPDF3.0 \cite{Ball:2014uwa}. 
Here, the theory results for $d\sigma/d\eta$ are all normalized 
to the data and for better visibility we do not include 
the uncertainties due to scale variations. The corresponding 
error band would partly lie outside the range of values shown 
in this figure and the data points are all found inside the 
theory error band (see Fig.~\ref{fig:1}). We can see in 
Fig.~\ref{fig:3} that the calculated cross sections 
do not depend strongly on the PDF input. This is, of 
course, not surprising since all PDF sets are based 
on fits to (almost) the same data. 

\begin{figure*}[b!]
\begin{center}
\includegraphics[width=7.8cm]{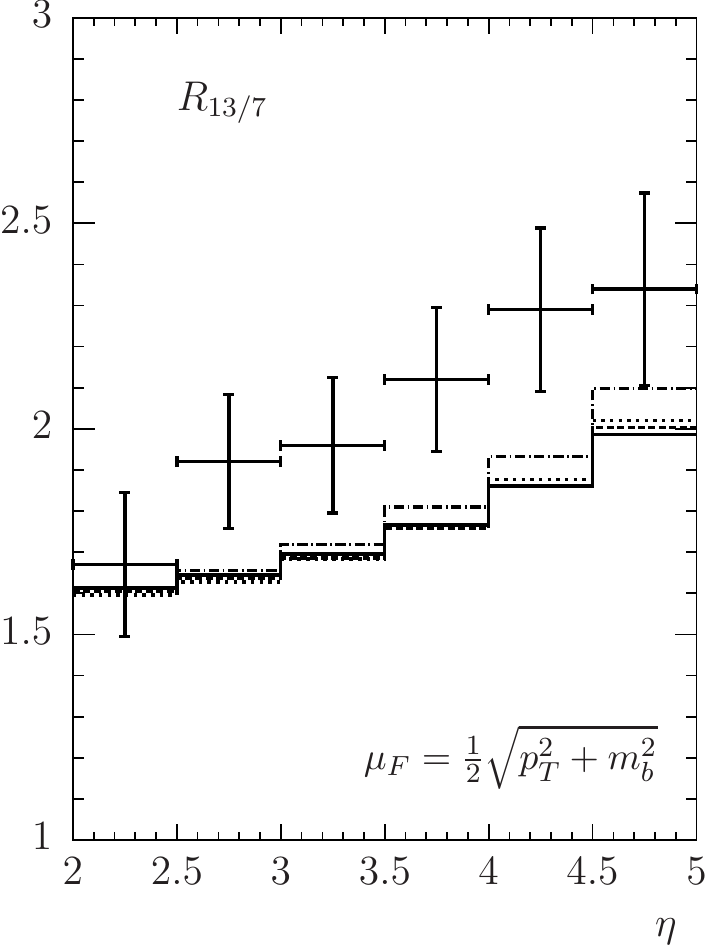}
\end{center}
\caption{
The ratio of the pseudorapidity distributions for $\sqrt{S} 
= 13$ and 7 TeV as in Fig.~\ref{fig:2} for different PDF sets. 
Full lines: CT14, dashed line: HERAPDF2.0, dotted line: 
NNPDF3.0, dash-dotted line MMHT. The full line agrees 
with the one in Fig.~\ref{fig:2}, but note the magnified 
scale in this figure. 
}
\label{fig:4} 
\end{figure*}

Figure~\ref{fig:4} shows the cross section ratio 
$R_{13/7}$ for the four different PDF sets. In the 
first few $\eta$ bins, the predictions are remarkably 
stable with respect to PDF variations. Only in the 
bin for the largest $\eta$ value, one can observe 
that the PDF set MMHT leads to a roughly 5\,\% increase 
of $R_{13/7}$ which brings the prediction closer to the 
measured value, but this shift is smaller than the 
experimental uncertainty. 

\begin{figure*}[b!]
\begin{center}
\includegraphics[width=7.0cm]{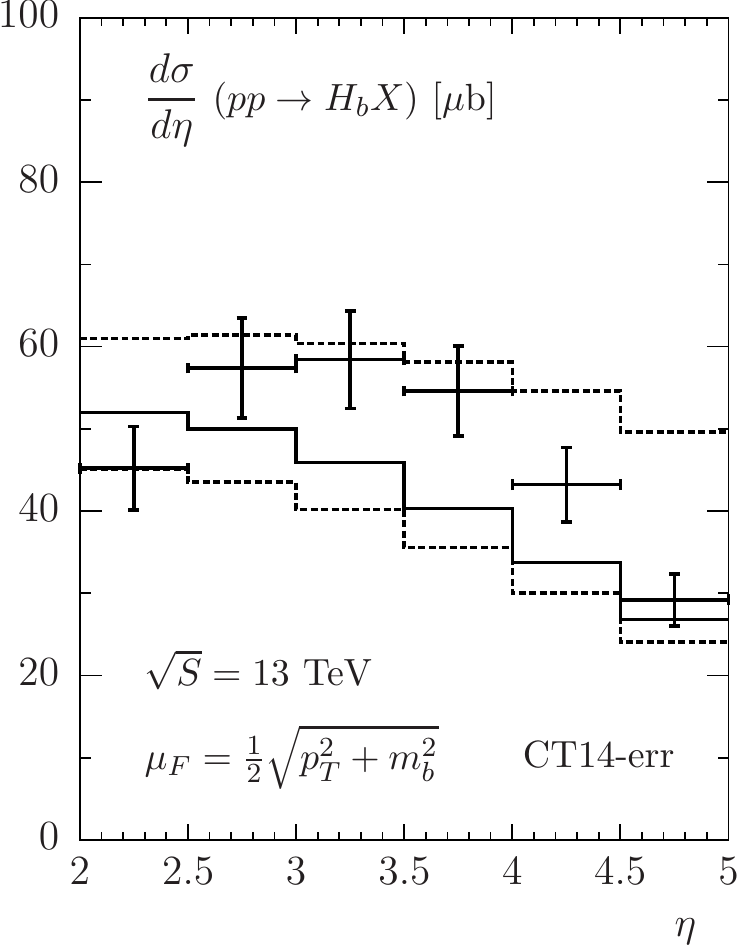}
~~
\includegraphics[width=6.8cm]{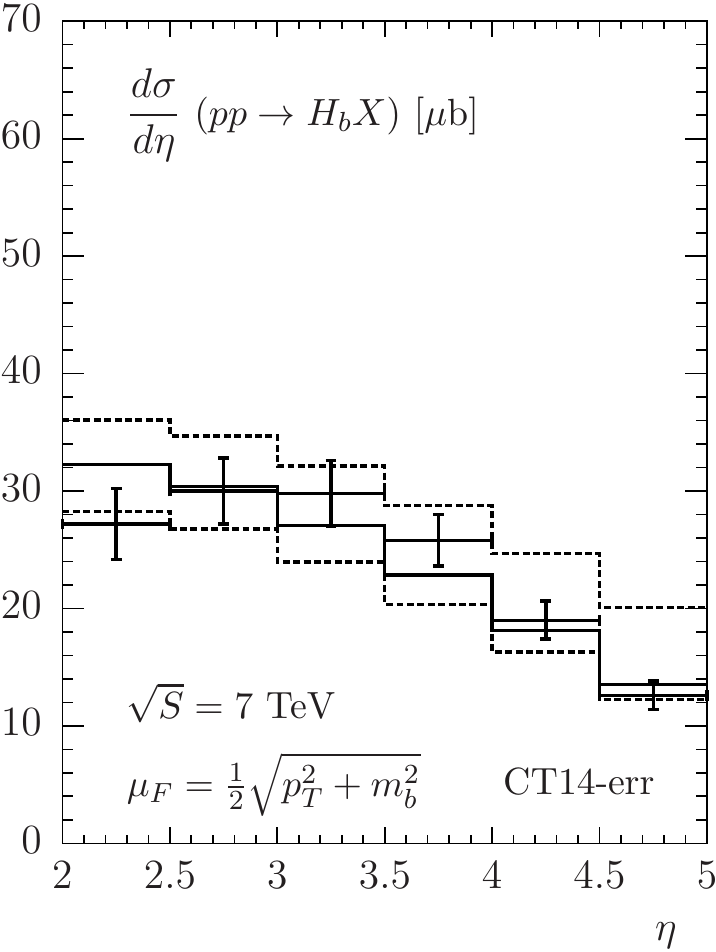}
\\
\includegraphics[width=7.0cm]{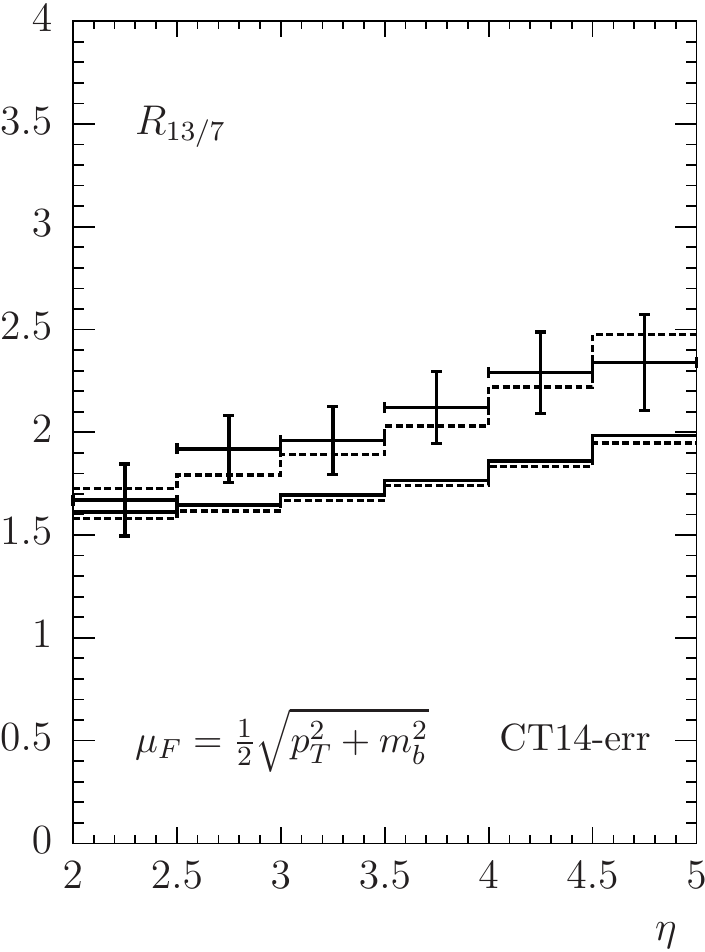}
\end{center}
\caption{
Envelope of the PDF uncertainty for the pseudorapidity 
distributions using all 28 member pairs of CT14 compared 
with LHCb data \cite{Aaij-Err:2016avz}. Upper left: 
$d\sigma/d\eta$ at $\sqrt{S} = 13$ TeV, upper right: 
$d\sigma/d\eta$ at $\sqrt{S} = 7$ TeV, lower: the ratio 
$R_{13/7}$. 
}
\label{fig:5} 
\end{figure*}

Assuming only ad-hoc changes of the proton PDF can not 
be expected to take full account of PDF uncertainties 
propagated from the data input on which the PDF fits 
are based. We therefore add results obtained by using 
uncertainty estimates which are now often also provided 
in a parametrized form by the PDF fitter collaborations. 
To be specific, we use the CT14 PDF set members with 28 
pairs of parameter eigenvalue variations. One pair is 
particularly interesting since it describes an enhanced 
or suppressed gluon distribution at very low $x$. Our 
results are shown in Fig.~\ref{fig:5} for $d\sigma/d\eta$ 
at $\sqrt{S} = 13$~TeV (upper left panel), $\sqrt{S} = 
7$~TeV (upper right panel), and for the ratio $R_{13/7}$ 
(lower panel). The dashed histograms represent the 
90\,\% confidence level\footnote{To obtain a 68\% CL 
   error band, one has to rescale the results by the factor 
   1/1.645.} maximal and minimal variations obtained 
from the CT14 eigenvalue variations, evaluated following 
the prescription described in Ref.~\cite{Lai:2010vv}. 
We find that indeed the PDF member pair 53 and 54 with 
an extreme choice of the gluon PDF has the strongest 
impact; the upward variation is dominated by one single 
eigenvector (member 53). The corresponding choice of the 
gluon distribution at low $x$ is possible since data still 
do not provide sufficient information. A similar observation 
has been made in a study of prompt neutrino fluxes from 
atmospheric charm production \cite{Benzke:2017yjn}. 
$d\sigma/d\eta$ depends strongly on the gluon PDF 
since $b$-hadron production is dominated by the 
sub-process $gg \to b \bar{b}$ at small $x \simeq 
10^{-4}$ and at small scales. This agrees with results 
of Ref.~\cite{Gauld:2017omh}. As can be seen in 
Fig.~\ref{fig:5}, the cross sections for $\sqrt{S} = 13$ 
and 7~TeV lie well inside the uncertainty band, while 
the ratio $R_{13/7}$ favors the upper limit of the 
CT14 error band. From our comparison we conclude that 
future PDF fits, in particular concerning the gluon PDF, 
could profit strongly from data for the $b$-quark 
production cross section ratio. 

\begin{figure*}[t!]
\begin{center}
\includegraphics[width=7.0cm]{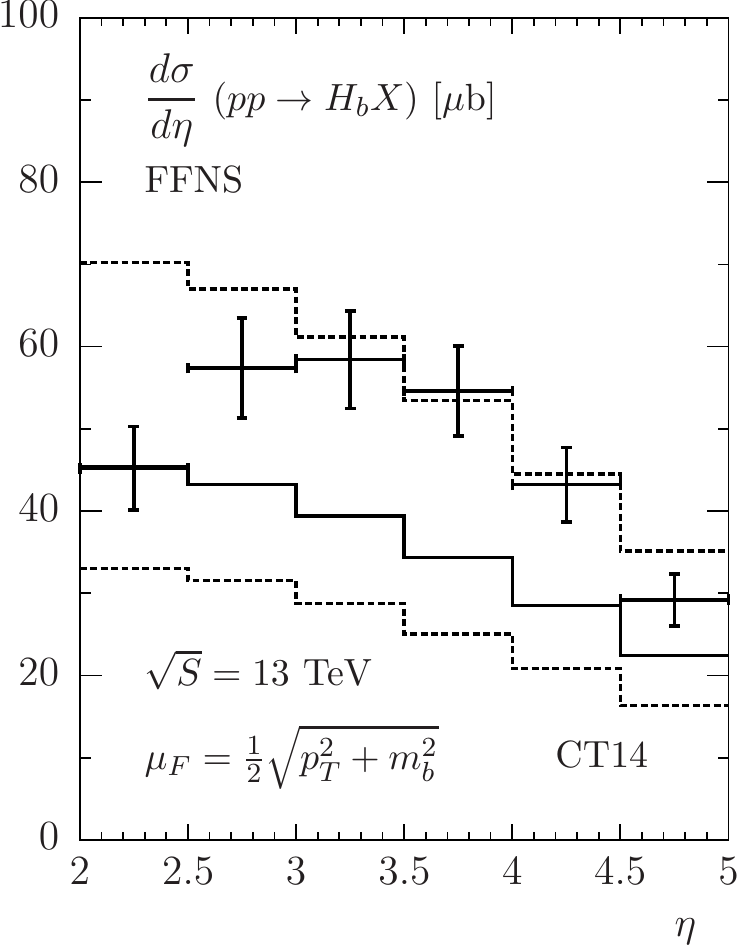}
~~
\includegraphics[width=6.8cm]{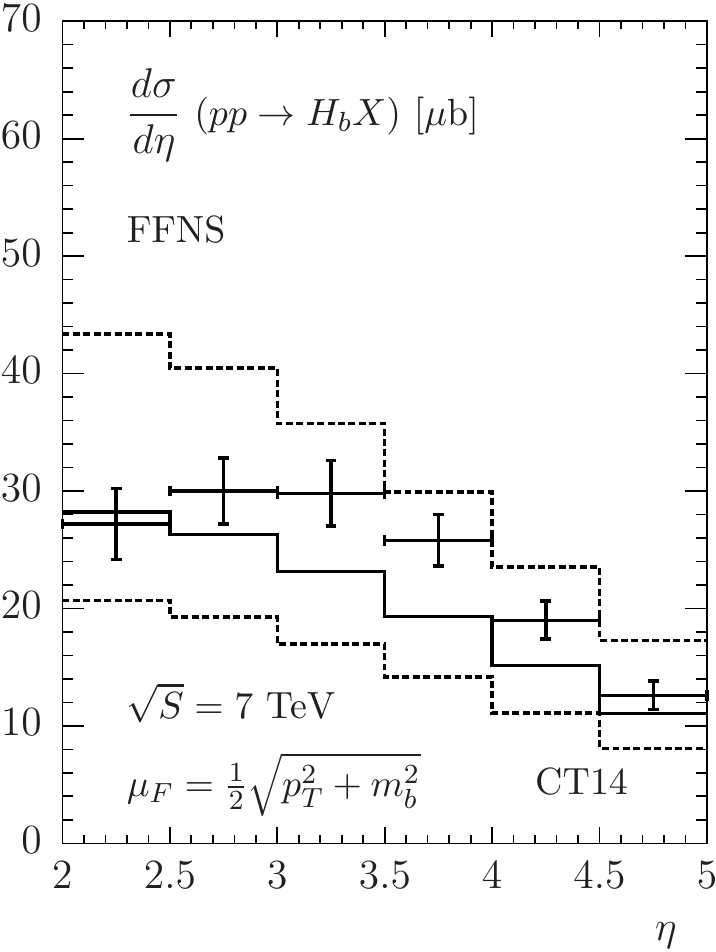}
\\
\includegraphics[width=7.0cm]{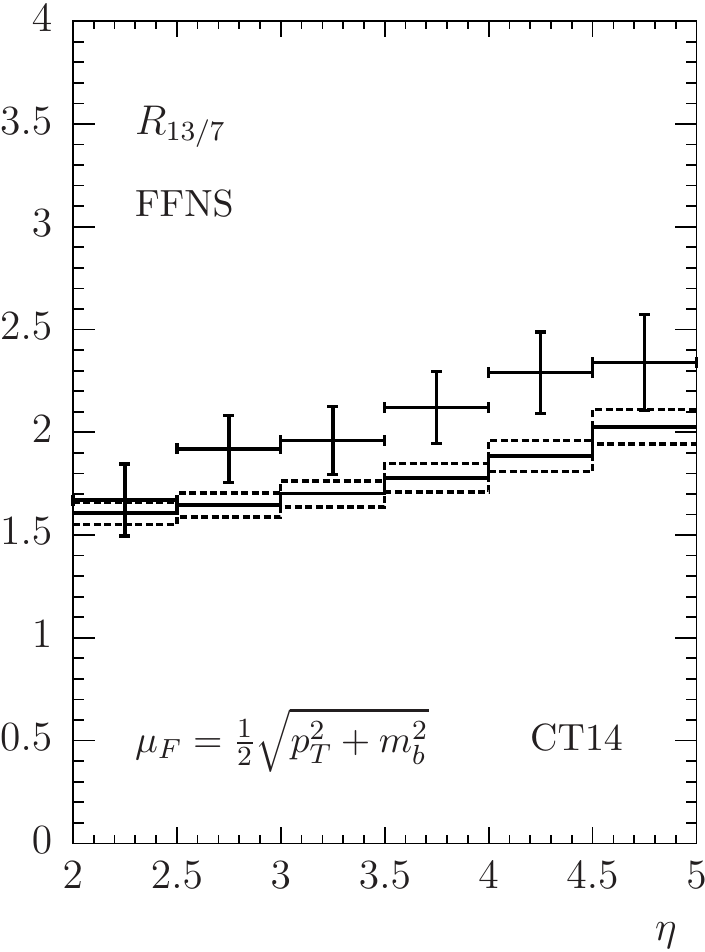}
\end{center}
\caption{
Pseudorapidity distribution for $b$-hadron production 
compared with LHCb data \cite{Aaij-Err:2016avz} as in 
Fig.~\ref{fig:1}, but for the FFNS. Upper left: 
$d\sigma/d\eta$ at $\sqrt{S} = 13$ TeV, upper right: 
$d\sigma/d\eta$ at $\sqrt{S} = 7$ TeV, lower: the ratio 
$R_{13/7}$. We have used CT14 PDFs and no FF. 
}
\label{fig:6} 
\end{figure*}

We finish this section with a comparison of the data with 
predictions obtained in the FFNS. In this approach, heavy 
quarks are produced at leading order only in the $gg$ 
channel and potentially large logarithms proportional 
to $\log (p_T/m_b)$ are not factorized into the PDFs and 
FFs. Consequently, there is no scale-dependent FF for the 
transition from $b$ quarks to $b$ hadrons. Since the LHCb 
data include all $b$ hadrons, we simply assume that the 
final state is given by a $b$ quark, i.e.\ we use a 
$\delta$ function as FF for the transition of a $b$-quark 
to a $b$ hadron. We use the CT14 PDFs including gluons 
and the light quarks $u$, $d$, $s$ and $c$ in the initial 
state, but no contribution from incoming $b$ quarks. The 
charm quark is treated as massless in this approach as 
before. In principle one should use a PDF set which is 
determined in the same framework of the FFNS with $n_f = 
4$, also including effects due to the non-zero charm mass 
\cite{Blumlein:2018jfm}, but from previous experience we 
do not expect large differences for such a more consistent 
approach. Results are shown in Fig.~\ref{fig:6}. The 
general picture looks quite similar to the GM-VFNS. The 
FFNS predictions are reduced by 15 to 20 percent, but 
the data are still within the theory uncertainty band. 
A comparison of the measured ratio $R_{13/7}$ with theory 
predictions in the FONNL framework \cite{Cacciari:2015fta} 
and using the NNPDF3.0 PDFs has also been presented in 
Ref.~\cite{Aaij-Err:2016avz}. The results look very similar 
to ours as shown in Figs.~\ref{fig:1} and \ref{fig:2}.

\clearpage

\section{Comparison with CMS data}
\label{Sec:compcms}

\begin{figure*}[b!]
\begin{center}
\includegraphics[width=7.26cm]{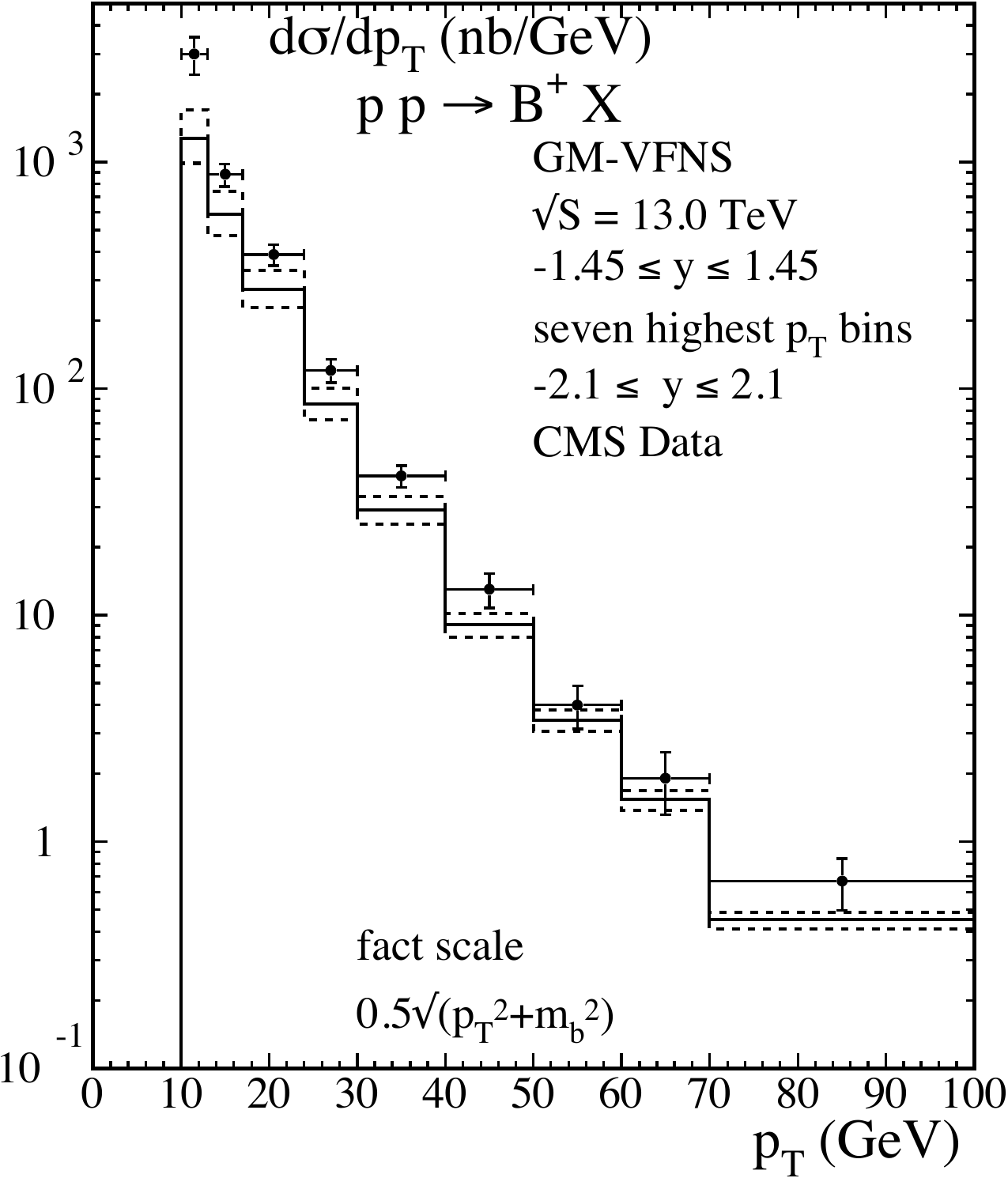}
~~
\includegraphics[width=7.1cm]{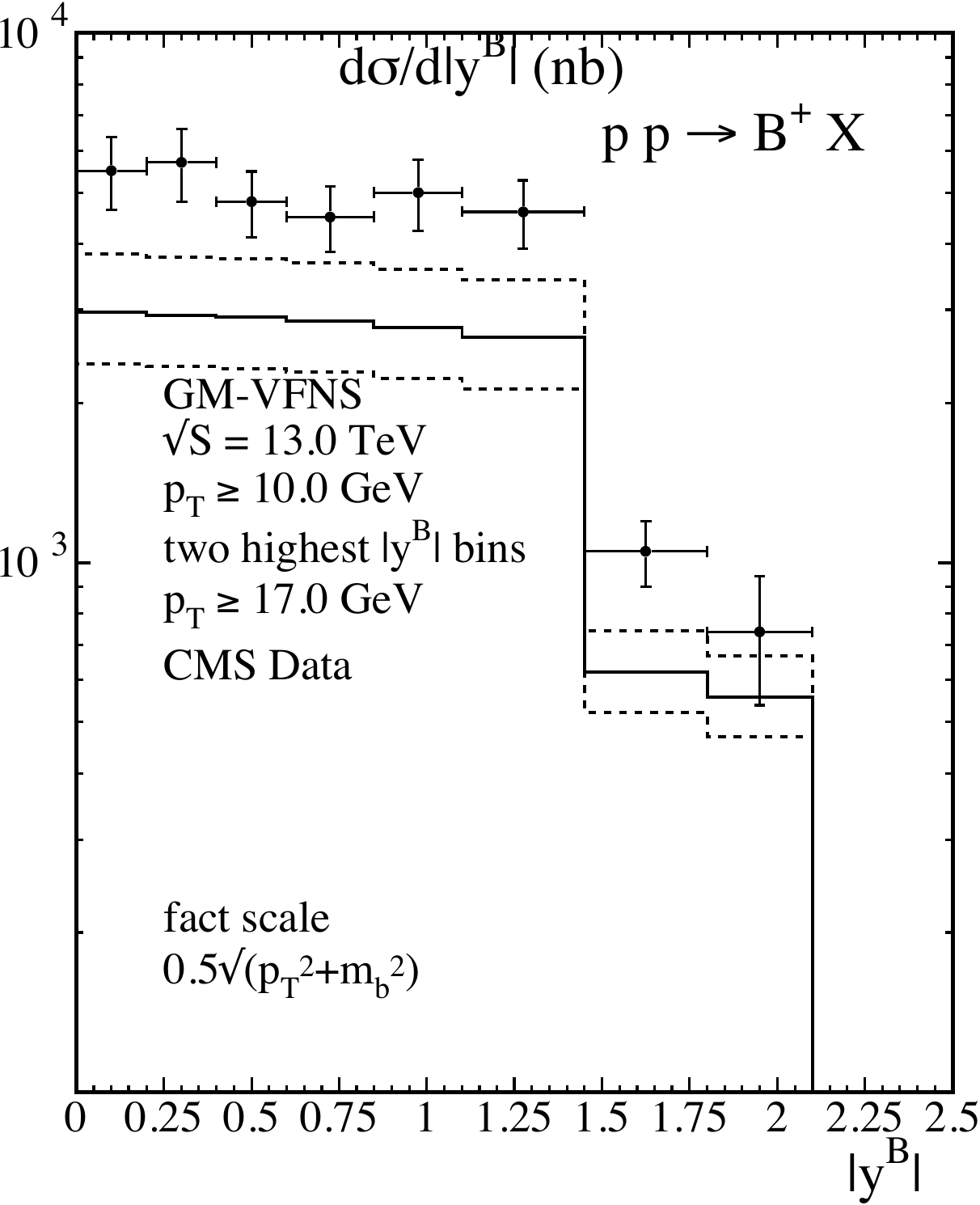}
~~
\end{center}
\caption{
$d\sigma/dp_T$ (left) and $d\sigma/d|y|$ (right) at $\sqrt{S} 
= 13$~TeV for $B^+$ production and comparison with data from 
the CMS collaboration \cite{Khachatryan:2016csy}. We have 
used $\mu_F = 0.5 \sqrt{p_T^2+m_b^2}$ and $\mu_R = 
\sqrt{p_T^2+m_b^2}$.
\label{fig:7} 
}
\end{figure*}

There are also data from the CMS collaboration for the 
ratio $R_{13/7}$ of cross sections for $b$-hadron production, 
however obtained from a measurement which includes only 
$B^+$-meson final states, $pp \rightarrow B^+ X$. Cross 
section data are available for $\sqrt{S} = 13$~TeV 
\cite{Khachatryan:2016csy} and $\sqrt{S}= 7$~TeV 
\cite{Chatrchyan:2011pw}. $B^+$ mesons were identified 
using the decay $B^+ \to J/\psi K^+$. The main difference 
to the LHCb analysis for $R_{13/7}$ is the kinematic range 
covered. CMS have used the rapidity $y$ of the reconstructed 
$B^+$ meson instead of the pseudorapidity as in the LHCb 
analysis. The ratio $R_{13/7}$ was obtained in the range 
$|y| < 1.45$ and $|y| < 2.1$, depending on different lower 
limits of the $p_T$ range: $p_T > 10$ GeV for $|y| < 1.45$ 
and $p_T > 17$ GeV for $|y| < 2.1$, respectively 
\cite{Khachatryan:2016csy}. 

We have calculated the cross section $d\sigma/dp_T$ at 
$\sqrt{S} = 13$~TeV for the nine $p_T$ bins in the region 
$10 < p_T < 100$ GeV as in the data analysis: two $p_T$ 
bins with $p_T \in [10,13]$ GeV and $p_T \in [13,17]$ GeV 
for $|y| < 1.45$ and seven $p_T$ bins for $p_T > 17$ GeV 
for $|y| < 2.1$. For consistency with the discussion in 
the previous section, we use also here the factorization 
scale $\mu_F = 0.5 \sqrt{p_T^2+m_b^2}$ and the renormalization 
scale $\mu_R = \sqrt{p_T^2+m_b^2}$ and vary $\mu_R$ up and 
down by a factor of 2. The result is shown in the left 
panel of Fig.~\ref{fig:7}. The agreement between our 
calculation and the measured cross section is satisfactory. 
The measured values for $d\sigma/dp_T$ lie mostly close to 
the results for the scale choice which leads to the maximal 
cross section, i.e. $\mu_R = 0.5 \sqrt{p_T^2+m_b^2}$. This 
relation between data and theory is similar to the comparison 
with FONLL predictions shown in Ref.~\cite{Khachatryan:2016csy}. 
Only the data point for the smallest $p_T$ bin lies well 
above the prediction and outside the theory error estimate. 

\begin{figure*}[b!]
\begin{center}
\includegraphics[width=7.0cm]{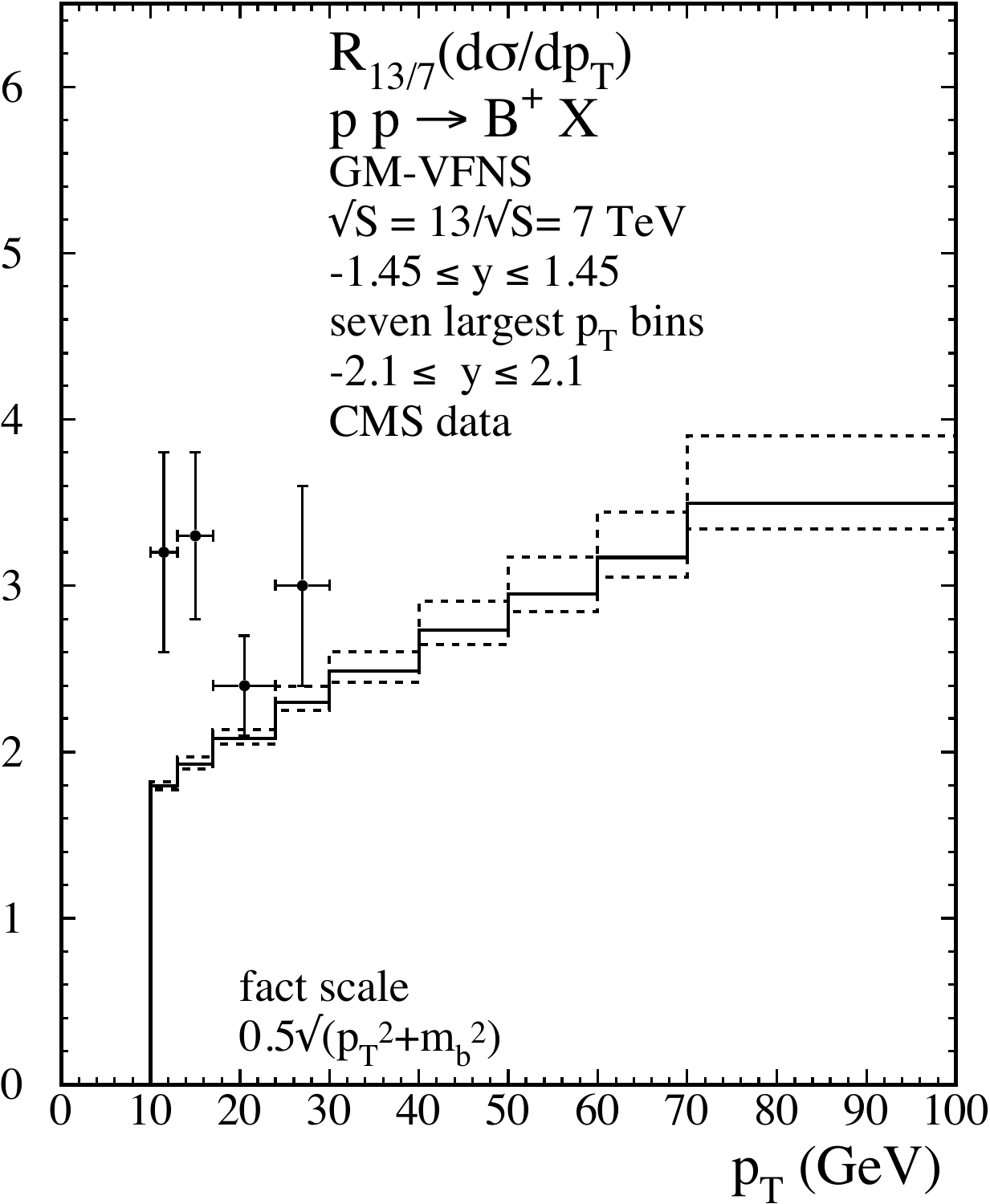}
~~
\includegraphics[width=7.0cm]{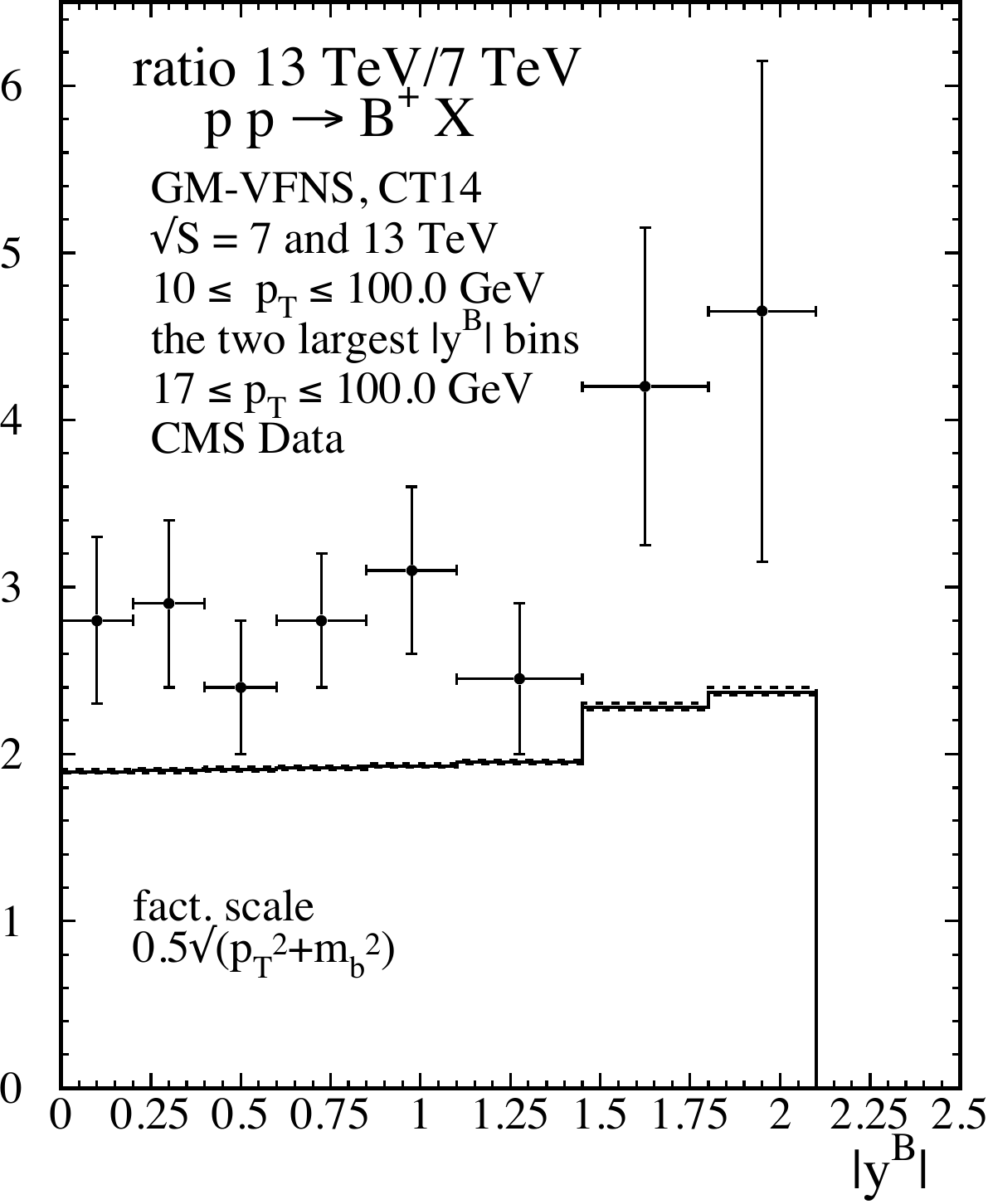}
\end{center}
\caption{
The ratio $R_{13/7}$ of cross sections for $B^+$ production 
and comparison with CMS data. The left panel is for the 
ratios of $d\sigma/dp_T$, the right panel of $d\sigma/d|y|$. 
The four data points in the left panel have been calculated 
from data given in \cite{Khachatryan:2016csy} and 
\cite{Khachatryan:2011mk}. We have used $\mu_F = 0.5 
\sqrt{p_T^2+m_b^2}$ and $\mu_R = \sqrt{p_T^2+m_b^2}$. 
\label{fig:8} 
}
\end{figure*}

In the right panel of Fig.~\ref{fig:7} we show the cross 
section $d\sigma/d|y|$ for eight $|y|$ bins integrated 
over the respective $p_T$ regions: $10 < p_T < 100$~GeV 
(first six lowest $|y|$ bins) and $17 < p_T < 100$~GeV 
(two highest $|y|$ bins). The data are compared to our 
predictions. The shape as a function of $|y|$ is very 
well reproduced, but the normalization of the cross section
data is $60\%$ higher than the calculated values. This is 
due to the fact that we have used the modified factorization 
scale $\mu_F = 0.5 \sqrt{p_T^2+m_b^2}$. With the original 
choice $\mu_F = \sqrt{p_T^2+m_b^2}$, which is more appropriate
for the cross section at larger $p_T$, we have found perfect 
agreement between data and theoretical predictions. 

\begin{figure*}[t!]
\begin{center}
\includegraphics[width=7.5cm]{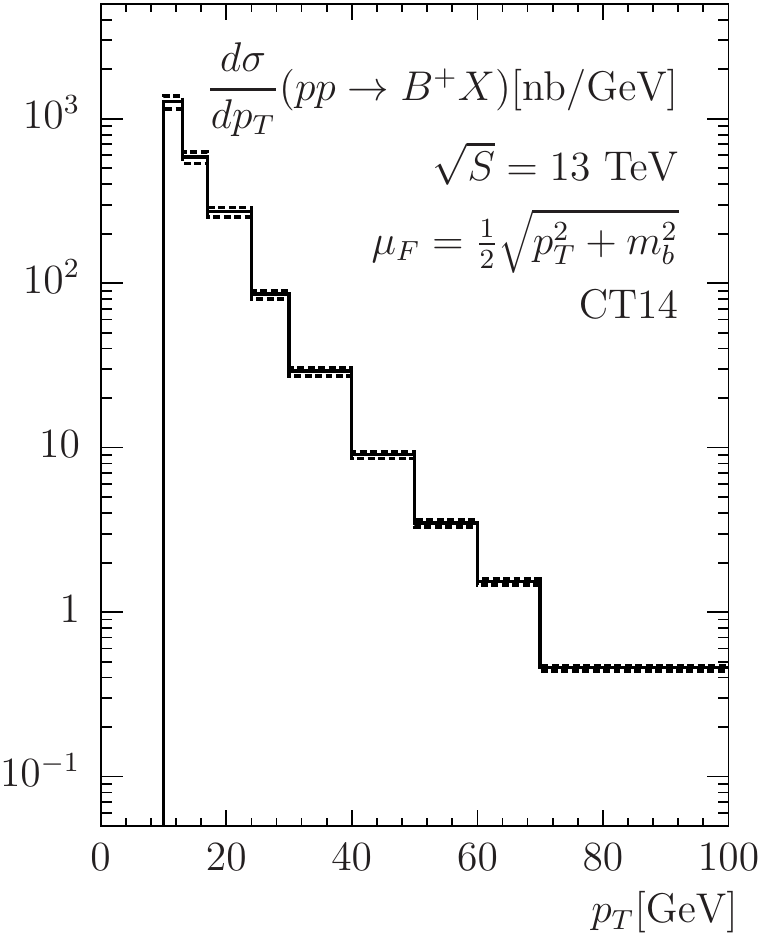}
~~
\includegraphics[width=7.0cm]{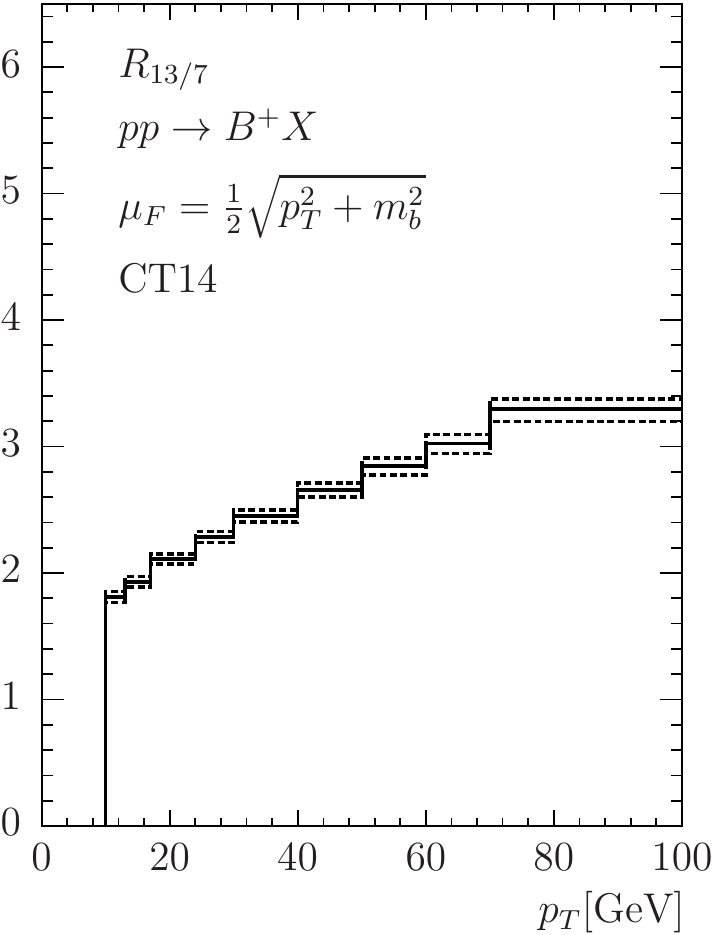}
\end{center}
\caption{
PDF uncertainties obtained from the 52 members of the 
CT14 PDF parametrization for $d\sigma/dp_T$ at $\sqrt{S} 
= 13$~TeV (left) and the ratio $R_{13/7}$ (right) for 
$B^+$ production at CMS.  
\label{fig:9} 
}
\end{figure*}

The ratios $R_{13/7}$ as a function of $p_T$ and $|y|$ 
are also reported in Ref.~\cite{Khachatryan:2016csy}. 
The data and the results of our calculation are shown in 
Fig.~\ref{fig:8} (left frame for the $p_T$-, right frame 
for the $|y|$-dependence). The four points for $p_T$ 
between $10$ and $30$ GeV are larger than our predictions, 
approximately by a factor of 1.6. Due to the rather large 
experimental errors, however, the disagreement in the CMS 
data is not very significant. The ratio $R_{13/7}$ as a 
function of $|y|$ is shown in Fig.~\ref{fig:8} (right 
frame). The measured ratio for eight $|y|$ bins, also 
taken from Ref.~\cite{Khachatryan:2016csy}, is again 
larger than the calculated ratio. Again, due to the 
large experimental uncertainties it is difficult to 
draw a definite conclusion from this comparison, 
similar to the case for the $p_T$-dependence of this 
ratio.  

In the large-$p_T$ region covered by the CMS measurements 
we do not expect any significant dependence of the theoretical 
cross sections on different choices of input for the proton 
PDFs. Here we decided to test PDF uncertainties using the 
26 pairs of eigenvalue variations of the CT14NLO PDFs 
\cite{Dulat:2015mca}. We present the results in 
Fig.~\ref{fig:9} for the cross section $d\sigma/dp_T$ 
for $pp \to B^+ X$ (left part) and for the ratio $R_{13/7}$ 
(right part). We find rather small uncertainties from 
this calculation, definitely much smaller than the 
uncertainties from scale variations for the theory 
prediction or from experimental sources for the data 
points. 

\clearpage

\section{Comparison with LHCb data for inclusive 
$J/\Psi$ production from $b$ mesons}
\label{Sec:compjpsi}

Another possibility to compare theoretical predictions 
with data for $b$-quark production with other final 
states is provided by a measurement of the LHCb collaboration 
of inclusive $J/\Psi$ production from $b$-meson decays. 
Data are available for the ratio $R_{13/8}$ as a function 
of $p_T$ and rapidity $y$ at $\sqrt{S} = 8$~TeV 
\cite{Aaij:2013yaa} and $\sqrt{S} = 13$~TeV 
\cite{Aaij:2015rla}. Some years ago, one of us 
together with P.~Bolzoni and B.~A.~Kniehl has considered 
this particular decay channel in the framework of the 
GM-VFNS incorporating theoretical input about the 
inclusive decay of $b$ hadrons into $J/\Psi$ mesons 
\cite{Bolzoni:2013tca}. Predictions have been compared 
with experimental data for $d\sigma/dp_T$ from the CDF, 
ALICE, ATLAS, CMS and LHCb collaborations existing at 
that time. In \cite{Bolzoni:2013tca} it was found that 
the data collected by these collaborations agreed 
reasonably well with the theoretical predictions. 

\begin{figure*}[t!]
\begin{center}
\includegraphics[width=7.0cm]{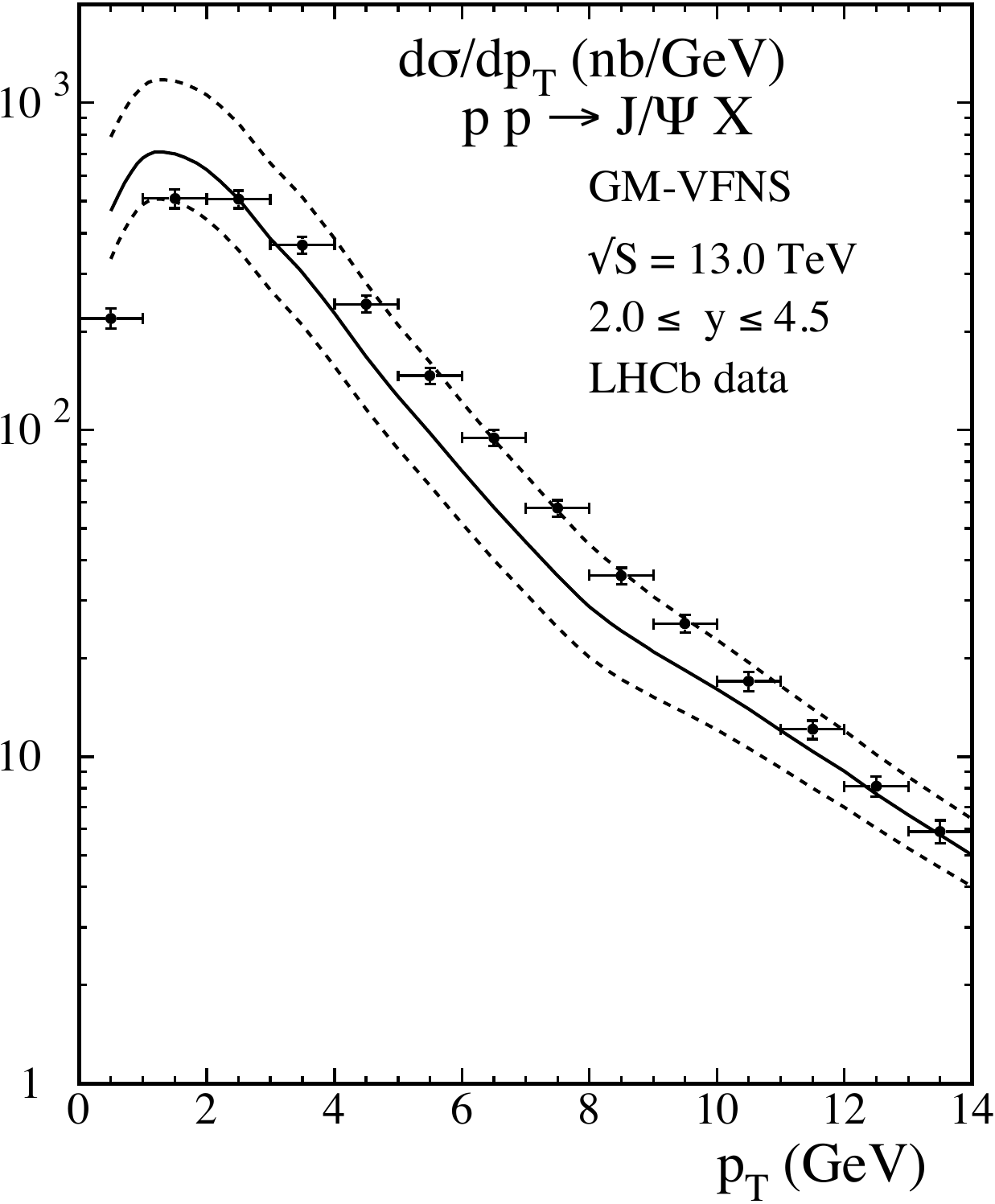}
~~
\includegraphics[width=7.0cm]{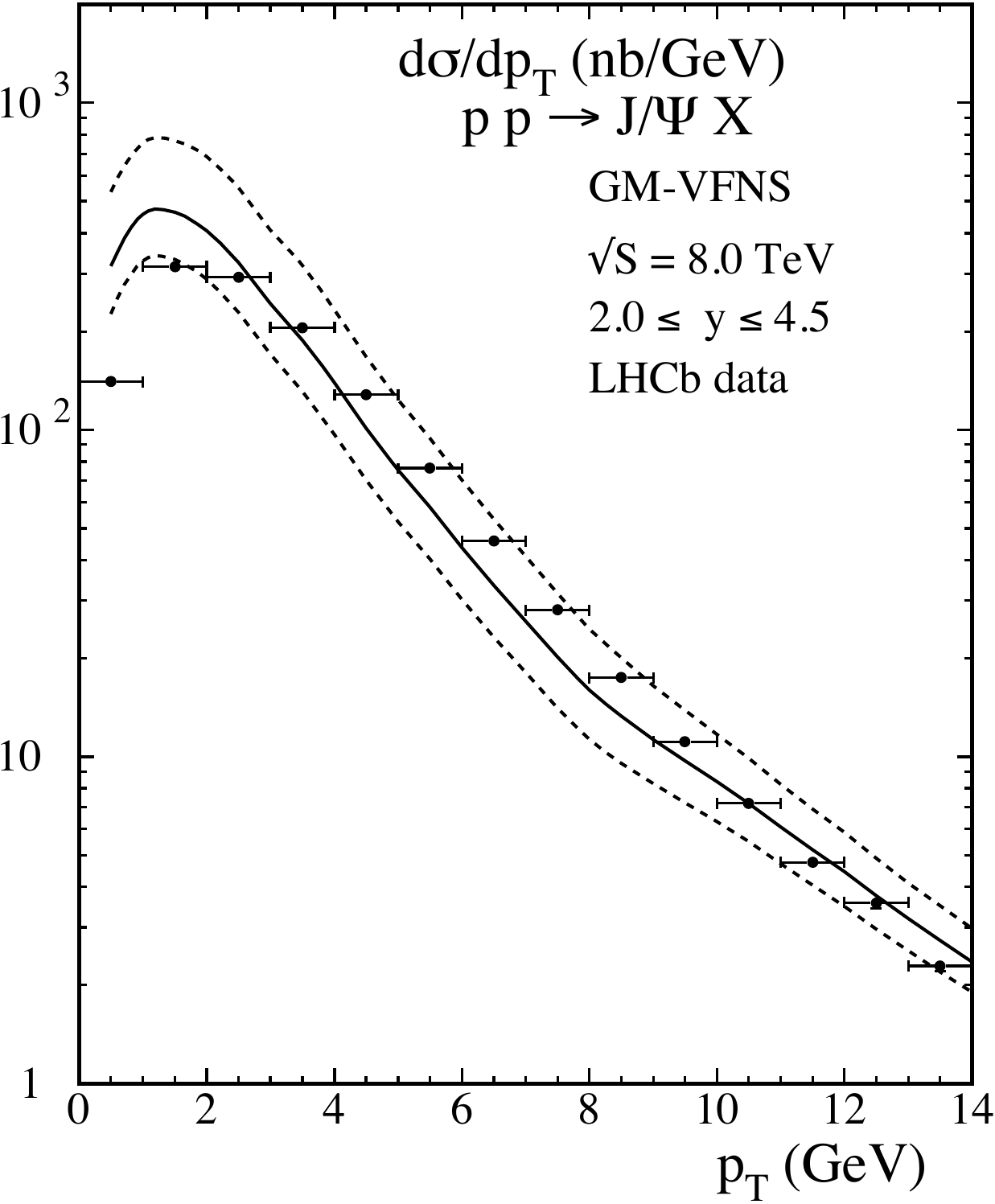}
\\[5mm]
\includegraphics[width=7.0cm]{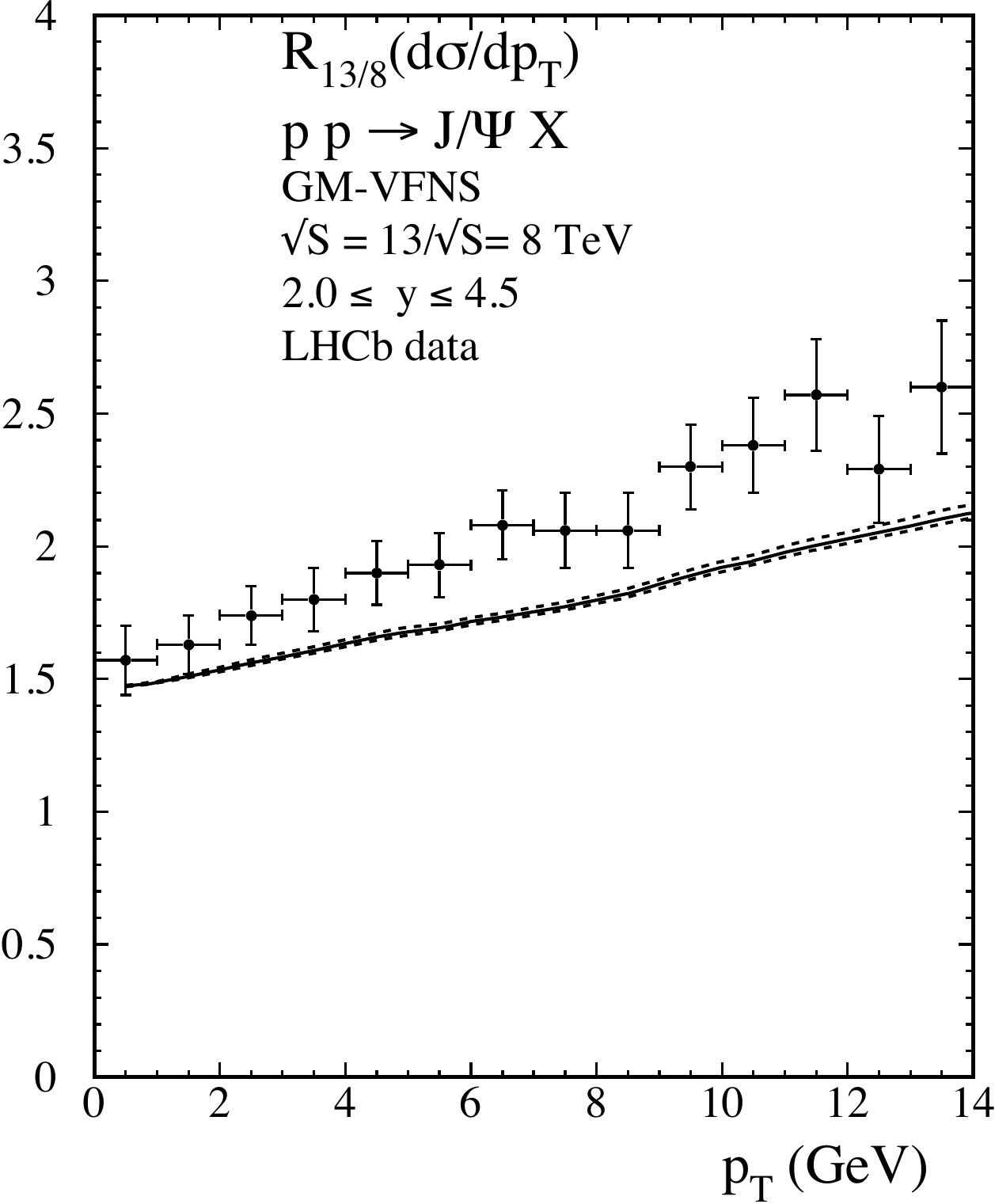}
\end{center}
\caption{
The cross section $d\sigma/dp_T$ for inclusive $J/\Psi$ 
production from $b$ mesons compared with LHCb data 
\cite{Aaij:2015rla} using CT14 at $\sqrt{S} = 13$ TeV 
(upper left), $\sqrt{S} = 8$ TeV (upper right), and 
their ratio (lower panel). 
\label{fig:10} 
}
\end{figure*}

Based on that earlier work we now calculate the 
transverse momentum and rapidity-dependent cross 
sections $d\sigma/dp_T$ and $d\sigma/dy$ for $\sqrt{S} 
= 8$~TeV and $\sqrt{S} = 13$~TeV and study the ratio 
$R_{13/8}$ as a function of $p_T$ and $y$ and compare 
with the experimental data presented in \cite{Aaij:2015rla}. 
In contrast to the earlier work on $J/\Psi$ production 
from $b$-hadron decay we now use the modified scale as 
described in Sec.~\ref{Sec:setup}. We expect that the 
scale choice $\mu_F = 0.5 \sqrt{p_T^2+m_b^2}$ is required 
to achieve a reasonable description of data also at low 
$p_T$. Our results for $d\sigma/dp_T$ as a function of 
$p_T$ for $\sqrt{S} = 13$~TeV and $\sqrt{S} = 8$~TeV are 
shown in Fig.~\ref{fig:10} in the left and right upper 
panels, resp. The theory results are compared with data 
taken from Refs.~\cite{Aaij:2015rla} (for $\sqrt{S} = 
13$~TeV) and \cite{Aaij:2013yaa} (for $\sqrt{S} = 8$~TeV). 
The experimental data points lie well inside the theoretical 
range which is determined from scale variations as usual. 
On average, the range between the default scale and the 
scale choice $\mu_R = \mu_F$ which leads to the maximal 
cross section is preferred by the data. The ratio 
$R_{13/8}$ for $d\sigma/dp_T$ as a function of $p_T$ is 
shown in the lower panel of Fig.~\ref{fig:10}. The 
theoretical prediction for $R_{13/8}$ increases from 
1.5 to about 2.2 and lies systematically below all data 
points, even when taking into account experimental 
uncertainties. The theoretical prediction for the ratio 
$R_{13/8}$ has essentially no scale dependence since we take 
the ratio of cross sections with identical scale parameters 
in the numerator and denominator. Numerical uncertainties 
are small and not shown in the figure.

\begin{figure*}[t!]
\begin{center}
\includegraphics[width=7.0cm]{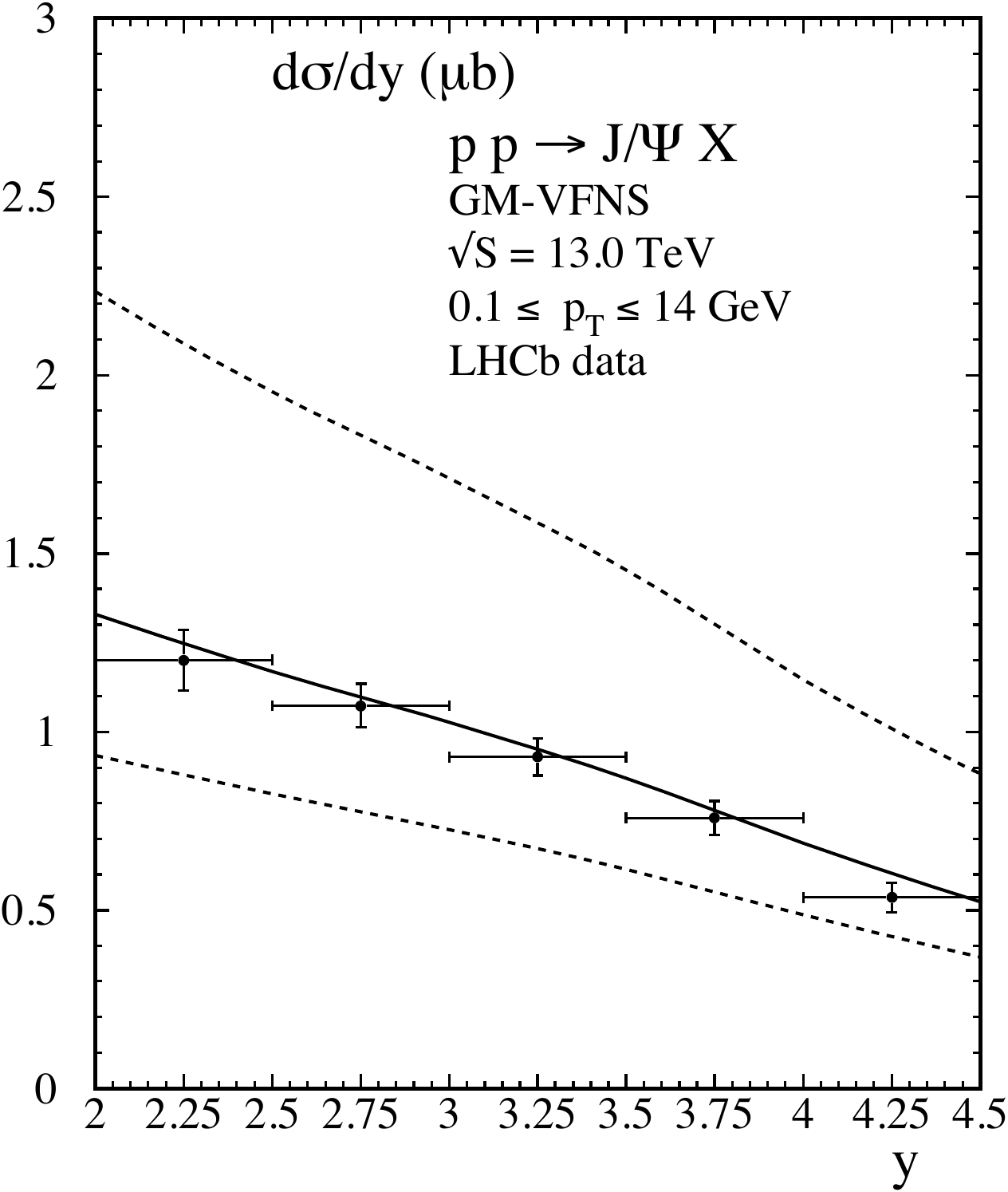}
~~
\includegraphics[width=7.0cm]{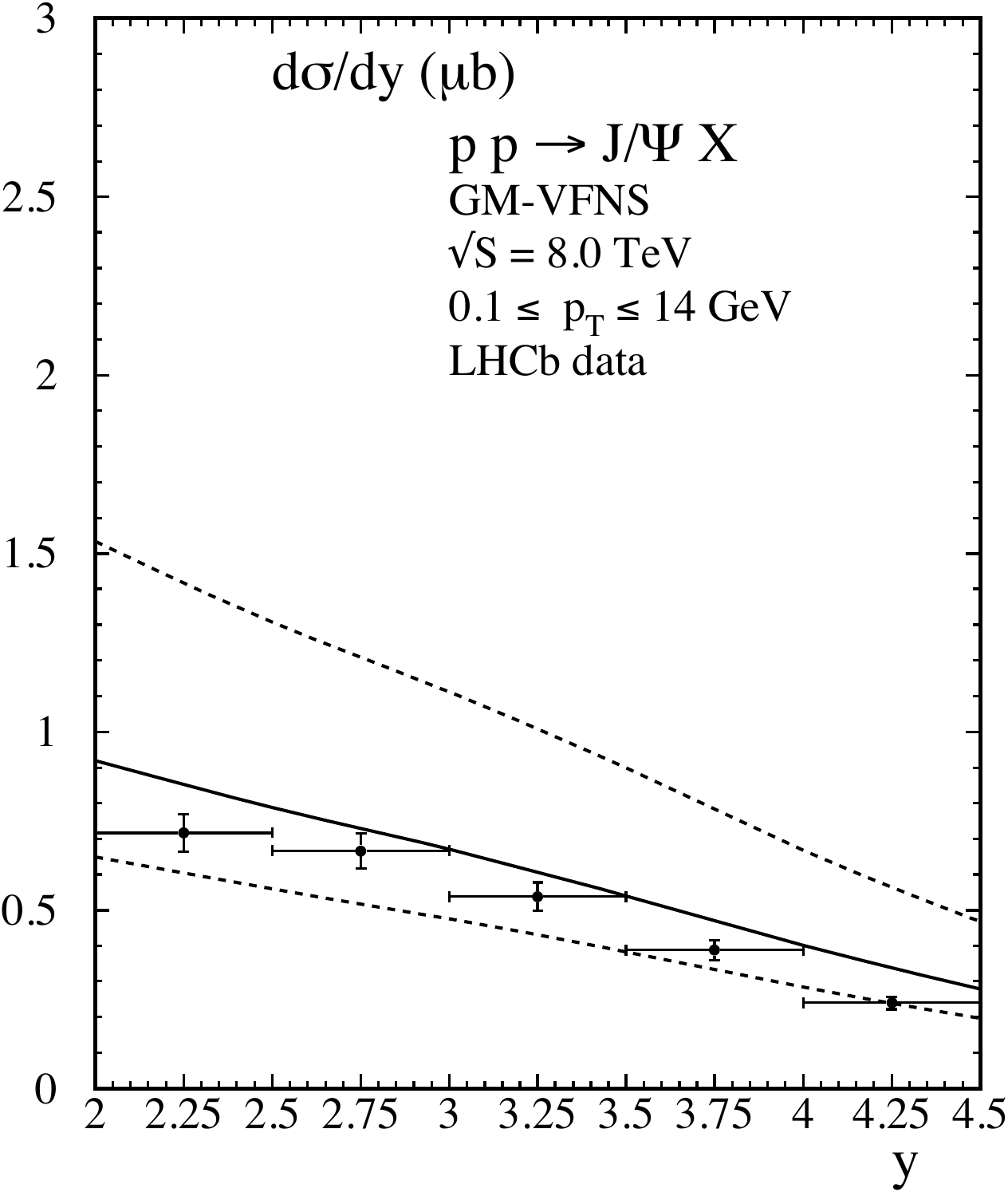}
\\[5mm]
\includegraphics[width=7.0cm]{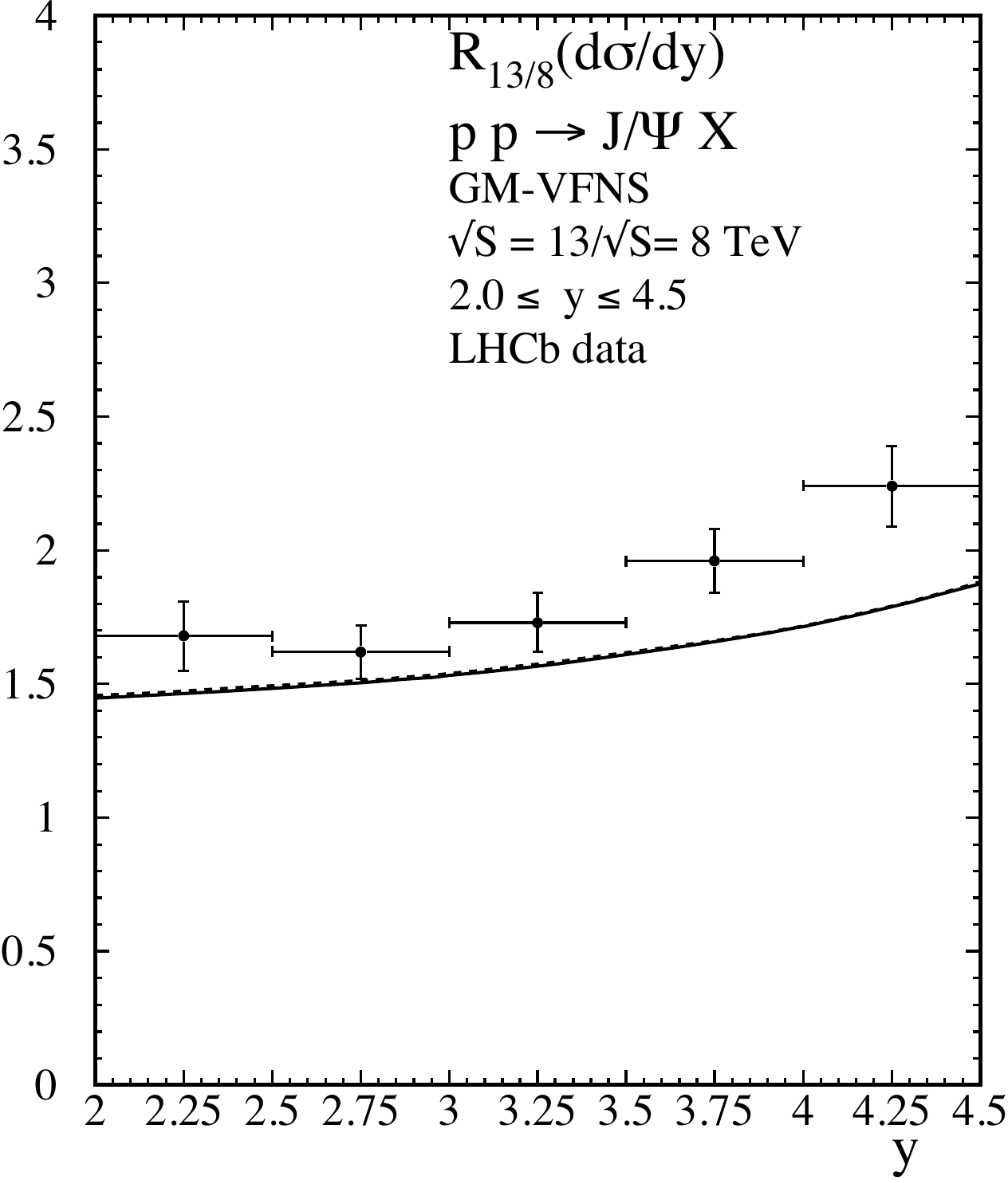}
\end{center}
\caption{
The cross section $d\sigma/dy$ for inclusive $J/\Psi$ 
production from $b$ mesons compared with LHCb data 
\cite{Aaij:2015rla} using CT14 at $\sqrt{S} = 13$ TeV 
(upper left), $\sqrt{S} = 8$ TeV (upper right), and 
their ratio (lower panel). 
\label{fig:11} 
}
\end{figure*}

The rapidity dependence of the cross section, $d\sigma/dy$, 
in the range $2.0 \leq y \leq 4.5$ integrated over the 
transverse momentum in the range $0 \leq p_T \leq 14$~GeV 
is presented in Fig.~\ref{fig:11}. The upper left panel 
shows the cross section as a function of $y$ for 
$\sqrt{S} = 13$~TeV, the upper right panel for $\sqrt{S} 
= 8$~TeV. Corresponding experimental data were given in 
Refs.~\cite{Aaij:2015rla,Aaij:2013yaa} in five bins 
of size $\Delta y = 0.5$. While the $13$~TeV data agree 
perfectly well with predictions for the default scale, 
the data at $\sqrt{S} = 8$~TeV lie somewhat below the 
prediction for the default scale, but still inside the 
theory uncertainty band whose lower limit is obtained 
with the choice $\mu_R = 4 \mu_F$. The theoretical 
result for the ratio $R_{13/8}$ as a function of $y$ 
is shown in the lower panel of Fig.~\ref{fig:11}. 
Here, the agreement between theory and data is rather 
marginal. A better agreement could be found if the 
prediction for $d\sigma/dy$ at $\sqrt{S} = 8$~TeV was 
slightly lower. In the present case, the strongest 
deviation between data and theory is in the last bins 
at the high $y$ values.  


\section{Conclusions}
\label{Sec:conclusion}

As a conclusion from our analysis of predictions for 
inclusive $b$-hadron production in $pp$ collisions 
at the LHC we can state that the majority of 
experimental data for the cross sections, differential 
in transverse momentum or in (pseudo-)rapidity, are 
reasonably well described by theory. This is mainly 
due to the large theory uncertainty due to variations 
of the renormalization scale. An exception is maybe 
seen in the comparison with $B^+$-meson production 
data from the CMS collaboration at $\sqrt{S} = 13$~TeV 
where the cross section data are somewhat higher than 
theory. 

In all the cases we found some tension between data 
and theory if the comparison is based on the cross 
section ratio for different center-of-mass energies. 
The data prefer slightly higher values of $R_{13/7}$ 
for $b$-hadron production measured by the LHCb 
collaboration and for $B^+$-meson production measured 
by the CMS collaboration. Also the LHCb data for the 
ratio $R_{13/8}$ of $J/\Psi$ production through decays 
from $b$ mesons is higher than our predictions. 

We found that the cross section ratios are remarkably 
stable with respect to variations of the renormalization 
scale if one follows the wide-spread assumption that 
the renormalization and factorization scales should 
depend only on the transverse momentum of the observed 
hadron. Our calculations show that a weakly 
$\sqrt{S}$-dependent choice of scales would further 
reduce the significance of the slight disagreement. 
Such a scale choice is theoretically neither 
well-motivated, nor completely unreasonable. 

As a consequence of the stability of theoretical 
predictions with respect to scale variations, one 
can also conclude that the cross section ratios 
will serve as important input for improved 
determinations of PDF parametrizations. This was 
exemplified in particular in a comparison with 
members of the CT14 PDF parametrizations. Future 
fits could result in an improved knowledge  
of the gluon PDF at low $x$ and, correlated 
with the low $p_T$ of the data, at low scale. In order 
to test such a possibility further it will be important 
to include also data at larger $p_T$. It would also be 
helpful if the LHCb collaboration could provide their 
data for $b$-hadron production with a higher value of 
the minimum transverse momentum. An extension of the 
kinematic range, both in transverse momentum and in 
(pseudo-)rapidity, is also important to test whether data 
from different experiments and with different $b$-hadron 
final states are compatible with each other.



\end{document}